\documentclass[lettersize,journal]{IEEEtran}
\usepackage{amsmath,amsfonts}
\usepackage{algorithmic}
\usepackage{algorithm}
\usepackage{array}
\usepackage[caption=false,font=normalsize,labelfont=sf,textfont=sf]{subfig}
\usepackage{textcomp}
\usepackage{stfloats}
\usepackage{url}
\usepackage{verbatim}
\usepackage{graphicx}
\usepackage{cite}
\usepackage{color}

\usepackage{amsmath,amsfonts}
\usepackage{algorithmic}
\usepackage{algorithm}
\usepackage{array}
\usepackage[caption=false,font=normalsize,labelfont=sf,textfont=sf]{subfig}
\usepackage{textcomp}
\usepackage{stfloats}
\usepackage{url}
\usepackage{verbatim}
\usepackage{graphicx}

\usepackage{cite}
\def\BibTeX{{\rm B\kern-.05em{\sc i\kern-.025em b}\kern-.08em
    T\kern-.1667em\lower.7ex\hbox{E}\kern-.125emX}}
\usepackage{balance}
\usepackage{url}
\usepackage{epstopdf}
\usepackage{amsmath}
\usepackage{amssymb}
\usepackage{amsthm}
\usepackage{booktabs}
\usepackage[table]{xcolor}
\usepackage{threeparttable} 
\usepackage{epsfig}

\usepackage{amsfonts}
\usepackage{multirow}
\usepackage{soul,color}

\usepackage[normalem]{ulem}
\useunder{\uline}{\ul}{}
\usepackage{color}
\usepackage{array}
\usepackage{acro}
\usepackage{listings}
\usepackage[underline=true]{pgf-umlsd}
\newcommand{\tabincell}[2]
{\begin{tabular}
		{@{}#1@{}}#2\end{tabular}}
\usepackage{setspace}

\DeclareRobustCommand{\acrodef}[2]{\DeclareAcronym{#1}{short=#1,long=#2}}
\acrodef{RRC}{Radio Resource Control}
\acrodef{ZMQ}{ZeroMQ}
\acrodef{LSTM}{Long Short-Term Memory}
\acrodef{LAL}{Listen-and-Learn}
\acrodef{SyAL}{Sync-and-Learn}
\acrodef{SoAL}{Source-and-Learn}

\acrodef{TS}{Technical Specifications}
\acrodef{NR}{New Radio}
\acrodef{URLLC}{Ultra-Reliable Low Latency Communications}
\acrodef{eMBB}{enhancing Mobile Broadband}
\acrodef{mMTC}{massive Machine Type Communications}
\acrodef{O-RAN}{Open Radio Access Network}
\acrodef{NAS}{Non-Access Stratum}
\acrodef{UE}{User Equipment}
\acrodef{MITM}{man-in-the-middle}
\acrodef{OTA}{over-the-air}
\acrodef{CN}{core network}
\acrodef{gNB}{gNodeB}
\acrodef{DoS}{Deny of Service}
\acrodef{BS}{Base Station}
\acrodef{RNTI}{Radio Network Temporary Identifier}
\acrodef{AR}{Augmented Reality}
\acrodef{IoT}{internet of things}
\acrodef{AI}{Artificial Intelligence}
\acrodef{SSD}{Systematic and Scalabile vulnerabilities
and unintended emergent behavior Detection Detection}
\acrodef{DCI}{Downlink Control Information}
\acrodef{CRC}{cyclic redundancy check}
\acrodef{ROC}{Receiver Operating Characteristics}
\acrodef{AUC}{Area Under the ROC Curve}
\acrodef{AKA}{Authentication and Key Agreement}
\acrodef{CDT}{cybersecurity digital twin}
\acrodef{NLP}{Natural Language Processing}

\newcommand\algorithmicprocedure{\textbf{procedure}}
\makeatletter
\newcommand\PROCEDURE[3][default]{%
  \ALC@it
  \algorithmicprocedure\ \textsc{#2}(#3)%
  \ALC@com{#1}%
  \begin{ALC@prc}%
}
\newcommand\ENDPROCEDURE{%
  \end{ALC@prc}%

}

\newenvironment{ALC@prc}{\begin{ALC@g}}{\end{ALC@g}}
\makeatother

\newcommand{\Comment}[1]{{\hskip3em //#1}}

\begin{document}

\title{Systematic Meets Unintended: Prior Knowledge Adaptive 5G Vulnerability Detection via Multi-Fuzzing}




\author{Jingda Yang,~\IEEEmembership{Student Member,~IEEE,}  Yanjun Pan,~\IEEEmembership{Member,~IEEE,} \\Tuyen~X.~Tran,~\IEEEmembership{Senior Member,~IEEE} and Ying Wang,~\IEEEmembership{Member,~IEEE,} \thanks{J. Yang and Y. Wang is with Stevens Institute of Technology. Y. Pan is with University of Arkansas and T. Tran is with AT\&T Labes Research. The corresponding
author is Y. Wang, ywang6@stevens.edu. This effort was sponsored by the Defense Advanced Research Project Agency (DARPA) under grant no. D22AP00144. The views and conclusions contained herein are those of the authors and should not be interpreted as necessarily representing the official policies or endorsements, either expressed or implied, of DARPA or the U.S. Government.}}


\markboth{Journal of \LaTeX\ Class Files}%
{Shell \MakeLowercase{\textit{et al.}}: A Sample Article Using IEEEtran.cls for IEEE Journals}

\IEEEpubid{0000--0000/00\$00.00~\copyright~2021 IEEE}

\maketitle

\begin{abstract} The virtualization and softwarization of 5G and NextG are critical enablers of the shift to flexibility, but they also present a potential attack surface for threats. However, current security research in communication systems focuses on specific aspects of security challenges and lacks a holistic perspective. To address this challenge, a novel systematic fuzzing approach is proposed to reveal, detect, and predict vulnerabilities with and without prior knowledge assumptions from attackers. It also serves as a digital twin platform for system testing and defense simulation pipeline. Three fuzzing strategies are proposed: Listen-and-Learn (LAL), Synchronize-and-Learn (SyAL), and Source-and-Learn (SoAL). The LAL strategy is a black-box fuzzing strategy used to discover vulnerabilities without prior protocol knowledge, while the SyAL strategy, also a black-box fuzzing method, targets vulnerabilities more accurately with attacker-accessible user information and a novel probability-based fuzzing approach. The white-box fuzzing strategy, SoAL, is then employed to identify and explain vulnerabilities through fuzzing of significant bits. Using the srsRAN 5G platform, the LAL strategy identifies 129 RRC connection vulnerabilities with an average detection duration of 0.072s. Leveraging the probability-based fuzzing algorithm, the SyAL strategy outperforms existing models in precision and recall, using significantly fewer fuzzing cases. SoAL detects three man-in-the-middle vulnerabilities stemming from 5G protocol vulnerabilities. The proposed solution is scalable to other open-source and commercial 5G platforms and protocols beyond RRC. Extensive experimental results demonstrate that the proposed solution is an effective and efficient approach to validate 5G security; meanwhile, it serves as real-time vulnerability detection and proactive defense.

\end{abstract}


\section{Introduction}\label{Background}
\IEEEPARstart{5}G \ac{NR} and NextG cellular networks promise a wide variety of heterogeneous use cases  that enable Ultra-Reliable Low Latency Communications (URLLC) and enhance Mobile Broadband (eMBB) and massive Machine Type Communications (mMTC) across various industries which require networking with unprecedented flexibility. Flexibility is key in 5G NR, providing performance enhancements and allowing vertical customization; however, these advances also increase the complexity of security in NR protocols and implementations \cite{Wang2021AI-PoweredOptimization}. While softwarization, virtualization, and disaggregation of networking functionalities as the key enablers of the needed shift to flexibility, it presents a potential attack surface to threats and requires rigorous testing against vulnerabilities, which are often computationally expensive and impractical in large-scale software stacks.

In existing large-scale codebases for both open source and commercially available 5G stacks, compared to the state-of-the-art security research in communication systems, which has resolved specific aspects or partitions of security challenges to achieve assurance, the 5G and NextG networks need to consider the security defense from a holistic perspective in detecting vulnerabilities and unintended emergent behaviors. In addition, defense strategies based on thorough testing often lack adaptation and robustness, facing variations of attacks and compromising 5G assurance. Furthermore, general system engineering approaches (e.g., system dynamics and agent-based modeling) are inadequate in describing both the qualitative and quantitative aspects of security features in 5G software stacks. Moreover, comprehensive cybersecurity assessments involving physical objects, particularly over critical infrastructures, can be expensive and time-consuming. \IEEEpubidadjcol

Unlike deterministic behaviors that can be verified via approaches like formal methods, detecting unintended emergent behavior in 5G software stacks requires repeatability and adds uncertainty to the results due to their stochastic nature and various use scenarios. Additionally, the recently adopted \ac{O-RAN} \cite{O-RANAlliance2018O-RAN:RAN}, characterized by machine learning algorithms, introduces less transparency to 5G communications. This uncertainty poses a significant challenge to traditional vulnerability detection methods, as they may not be able to effectively identify vulnerabilities arising from unexpected inputs or behaviors resulting from machine learning algorithms. Therefore, an efficient and systematic scheme that is based on experimental results for detecting vulnerabilities and unintended emergent behaviors is crucial for ensuring the security and robustness of 5G systems. Experimental work in the context of 5G shifts simulation-driven research used in previous mobile network generations to system implementation prototyping\cite{Wang2021DevelopmentResearch}\cite{Breen2021POWDER:Research}\cite{Wang2022FromSystem}. This change stems from several factors including the widespread adoption of programmable Software Defined Radios (SDR), network function virtualization (NFV), and subsequent open-source softwarization of mobile network functions through various projects such as Open Air Interface (OAI) \cite{InformationstechnischeGesellschaftimVDEWSAAntennas.},   srsRAN \cite{Gomez-Miguelez2016SrsLTE:Experimentation}. The change  also enables digital twins to emerge as a disruptive concept for testing complex, large-scale 5G network stacks with limited resources. A digital twin enables systems analysis, design, optimization, and evolution to take place fully digitally or in conjunction with a cyber-physical system. Compared to traditional engineering approaches, digital twin solutions offer enhanced speed, accuracy, and efficiency~\cite{holmes2021digital}. The applications of digital twin range from well-defined low-flow device control communication, such as in Industry 4.0~\cite{teisserenc2021adoption}, to more sophisticated applications involving large volumes of data flow fields, such as Augmented Reality (AR)~\cite{bohm2021augmented}.

The concept of digital twins has been applied to cybersecurity and risk management in communication systems. Nguyen et al. \cite{nguyen2021digital} proposed a systematic approach to using digital twins for developing and deploying complex 5G environments and risk prevention systems. Additionally, Jagannath et al. \cite{jagannath2022digital} developed an AI cloud modeling system to mitigate risks associated with developing innovative technologies on existing systems. Network digital twins, in conjunction with traditional fuzzing approaches, offer comprehensive feasibility proof and evaluation for development and deployment on physical systems~\cite{almasan2022network}. Furthermore, digital twins technology can be utilized to formulate specific and efficient standards for security training \cite{vakaruk2021digital}, extending beyond software design and development.

Though various digital twin applications exist in manufacturing and research, the exponential increase in data volume and volatile environments presents significant challenges for using digital twins in physical systems. For example, simulating and identifying unintended emergent behaviors in 5G to provide scalable cybersecurity assurance through digital twins remains difficult. As a result, digital twins are often limited to descriptive rather than actionable functions in 5G and cybersecurity fields \cite{Nguyen2021DigitalBeyond}. This study aims to develop a digital twin platform that is not only descriptive but also actionable against potential or actual attacks in a physical 5G system, from the micro-atomic to the macro-geometric level. In addition, effective fuzzing strategies are developed on the platform to detect vulnerabilities and unintended emergent behaviors in 5G specifications and implementations.

In addition to an efficient testbed, many studies have proposed research on 5G protocol vulnerability detection and the extension to a critical area application. For instance, 3GPP TS 33.501a described how a significant number of pre-authentication messages are sent in an unencrypted format, which can be exploited to launch DoS attacks and obtain sensitive information, such as the location of mobile subscribers in 5G/LTE \cite{Jover2019SecuritySpecifications}. In \cite{8403769}, the authors identified the weakest links and channels vulnerable to sniffing and spoofing in the 5G NR framework. Hussain et al. \cite{Hussain20195Greasoner:Protocol} proposed a property-directed approach for qualitative and quantifiable analysis. Innovative strategies such as the grammar-based fuzzing approach with a Markov chain model have recently been proposed to generate high-risk fuzzing inputs~\cite{al2021grammar}. Similarly, other stateful transition models have been introduced to efficiently locate vulnerabilities~\cite{yang20235g,wang2022automated,he2022intelligent}. In an effort to further refine the fuzzing scope, formal verification has been incorporated into fuzzing strategies as demonstrated by HyPFuzz~\cite{chen2023hypfuzz}. Capitalizing on advancements in deep learning technologies, Rainfuzz~\cite{rullo2021rainfuzz} employs reinforcement learning to generate a heatmap, facilitating an estimation of risk associated with varying permutations of fuzzing cases. Additionally, Natural Language Processing (NLP) has been introduced to analyze vulnerabilities directly from the source code~\cite{singh2022cyber}. In a bid to enhance vulnerability assessment, the development of security metrics~\cite{moukahal2021vulnerability} and dependent fields~\cite{salazar2022formal} offers a more comprehensive visualization of vulnerability evaluation. These developments continue to contribute to the effectiveness and efficiency of vulnerability detection and risk assessment.
Despite of the substantial contribution to protocol-based vulnerability detection, a comprehensive and systematic approach for detecting vulnerabilities and unintended emergent behaviors in the entire protocol, considering varying perspectives on prior knowledge and fuzzing levels, remains unaddressed.

Among vulnerability detection approaches, fuzz testing has been extensively used in large-scale 5G and beyond systems for cybersecurity purposes. 
Nevertheless, the major challenge in this area remains computational complexity, which tends to increase exponentially with the size of the protocol complexity. To confine the detection within the protocol scope, Potnuru et al.~\cite{Potnuru2021Berserker:5G} proposed a protocol-based fuzz testing to generate fuzz testing cases with all possible identifiers and provided a comprehensive understanding of how the reaction to different protocol-based attacks. Han et al. proposed mutation-based fuzz testing~\cite{han2012mutation}, which can generate extreme cases, like buffer overflow or incorrect format. Combining the advantages of protocol and mutation, Salazar et al. provided a rule-based fuzzer~\cite{Salazar20215Greplay:Injection}, which can cover all protocol-based cases and part of extreme cases. In~\cite{ma2017semi}, Ma et al. proposed a state transaction method to analyze serial attacks, which can be achieved by modifying different messages in different states. Their approaches significantly augmented the complexity and diversity of attacks. However, assuming prior knowledge may implicitly limit the applicability of vulnerability detection methods, as attackers will exploit vulnerabilities based on the most efficient means available, utilizing their knowledge of the system. Successful tokenized-based general-purpose fuzzers (GPF), such as LZfuzz~\cite{bratus2008lzfuzz}, eliminate the requirements for access to well-documented protocols and implementations, while focusing on plain-text fuzzing. Additionally, the non-selective traverse type of fuzzing relies on massive computation resources~\cite{ma2017semi}. To address the challenges of the prior knowledge requirements and computation complexity, we propose a multiple dimension multi-layer protocol-independent fuzzing framework based on digital twin system, which is combined with machine learning algorithms aiming to detect protocol vulnerabilities and unintended emergent behaviors in fast-evolving 5G and NextG specifications and large-scale open programmable 5G stacks.

\subsection{Contributions}
The Dolev-Yao formal verification model \cite{dolev1983security} has been widely used in the field of cryptography and security protocol analysis. It simplifies the analysis of cryptographic protocols by assuming that an attacker has the ability to intercept, modify, and fabricate messages. The lack of consideration of specific algorithmic weaknesses limits its application in large-scale distributed networks \cite{syverson2000dolev}. The prior knowledge required by attackers can be considered as the cost of the attack and the impacts or risks to the target network can be considered as the gain. Counting on the cost and gain for attacks being performed in unexpected scenarios, it is necessary to build a systematic solution that adapts to scenarios and detects vulnerabilities and their impacts. Thus, based on the awareness of prior knowledge, our fuzzing strategies include `LAL', `SyAL' and `SoAL', which are analogous to black-box, grey-box, and white-box models. Prior knowledge, in this context, refers to any information that an attacker possesses about a system or element before attempting to exploit vulnerabilities. This information includes protocols, synchronization information, zero-day exploits, or any other relevant information that can be used to exploit potential vulnerabilities. Furthermore, we develop a digital twin framework that connects to an experimental platform to enable systematic and accumulative vulnerability detection for 5G and NextG protocols through fuzz testing, leveraging our existing work in \cite{JingdaYang20235GListen-and-Learn} and  \cite{Dauphinais2023AutomatedSystems}.  The RRC protocols and implementations on srsRAN are adapted to serve as digital twin proof-of-concept of the designed system, and a relay model acts as an attacker.


In particular, the proposed \ac{LAL} command-level strategy assumes no prior knowledge of protocols nor access to code stacks. The protocol-independent characteristic of it enables an automatic verification for 5G and NextG protocols and large-scale open programmable stacks release.
Then by leveraging some prior and domain knowledge, a more strategic grey-box approach, \ac{SyAL} command-level fuzzing, is formed to achieve higher efficiency and accuracy in detecting vulnerabilities. 
In scenarios with access to the source code, the designed \ac{SoAL} bit-level fuzzing works as a white-box strategy to perform a more sophisticated approach on the identified high-risk commands via \ac{LAL} and \ac{SyAL}.  Our proposed fuzzing system offers sufficient automation and efficiency that could serve as a feasible approach to validate security for 5G protocols and implementations. It also enables real-time system vulnerability detection and proactive defense.

\begin{figure}[!t]
\centering\includegraphics[width=0.5\textwidth]{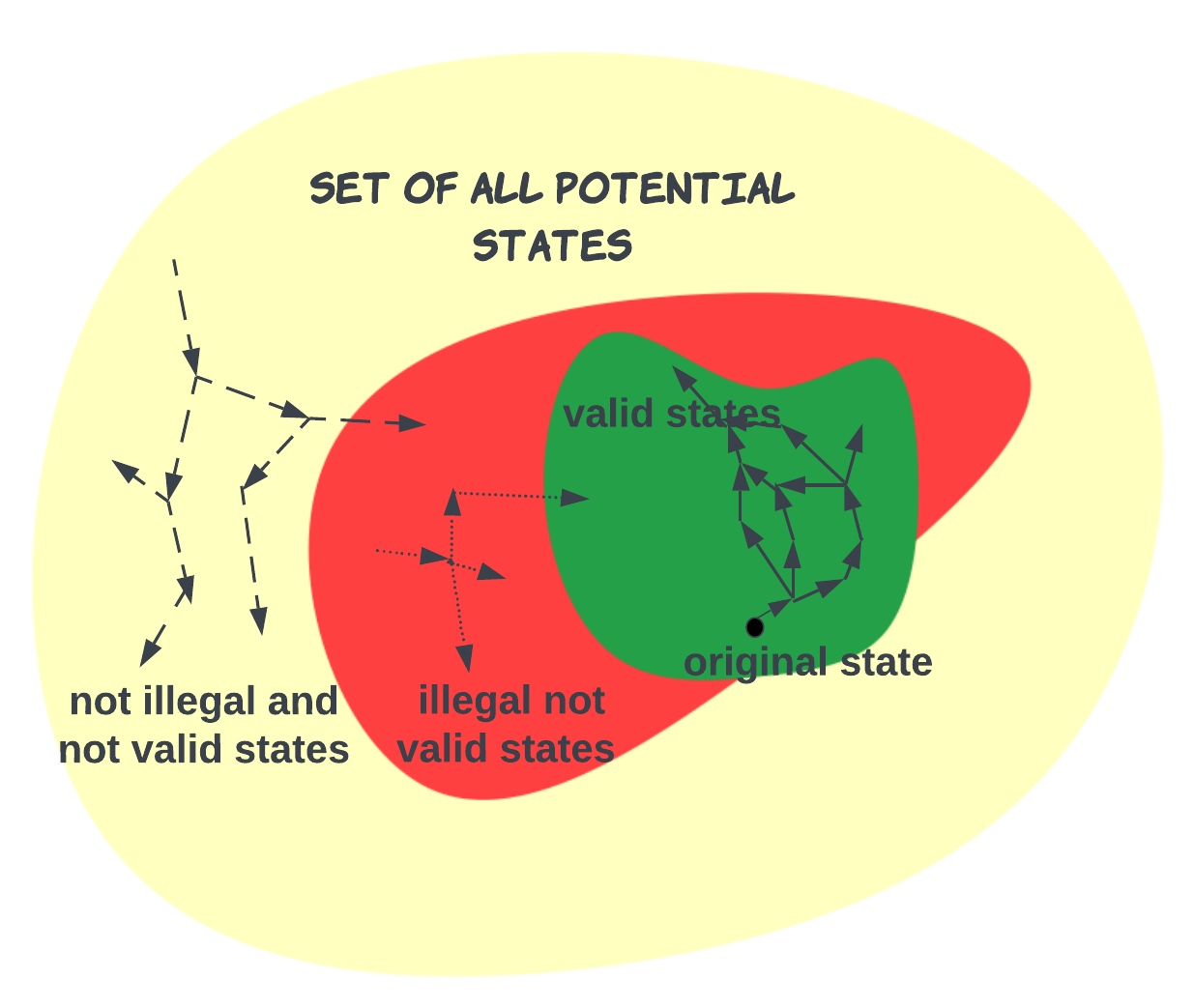}
    \caption{Definition of fuzz testing region.}
    \label{fig:states}
\end{figure}

The states of commands in wireless connections can primarily be divided into three zones: valid and legal states (green zone), illegal and invalid states (red zone), and not illegal and not valid states (yellow zone), as illustrated in Fig. \mbox{\ref{fig:states}}. `Legal' indicates whether a command can pass the integrity check, while `valid' refers to whether the command can function as intended. For example, states in the not illegal and not valid zone are those that do not trigger the defense mechanism but can introduce potential threats. Compared to intended attacks, which are defined in the protocol and located in the red zone, unintended vulnerabilities are more challenging to detect in 5G wireless communication. Therefore, our focus is on detecting unintended vulnerabilities, which are difficult to identify using the integrity protection mechanisms of the \ac{UE} and \ac{gNB}.

Our main contributions are summarized below:
\begin{enumerate}

    \item Taking into account the attacker strategies with different levels of prior knowledge, we design three fuzzing strategies named \ac{LAL}, \ac{SyAL}, and \ac{SoAL}. These strategies offer an efficient and comprehensive solution for vulnerability detection in 5G specifications and stacks.

    \item We propose a probability-based fuzzing approach that can reduce the average number of fuzzing cases expected to detect a vulnerability from linear to logarithmic growth, resulting in significant scalability and efficiency improvements for complex systems.

    \item We design a renovated 5G \ac{CDT} platform based on classical 5G cybersecurity modeling. Compared to existing ones, the introduced platform is not only describable but also actionable for potential or actual attacks in a physical 5G system.
    
    \item A proof-of-concept of the designed framework piloting \ac{RRC} protocols in the srsRAN platform is developed. The discovered vulnerable states and transactions of the \ac{RRC} protocol provide insights for fortifying 5G and NextG protocols.

    \item The digital twin solution can directly scale to other existing and future open-source and commercial 5G platforms and protocols other than RRC.

    
\end{enumerate} 

The rest of the paper is organized as follows: 
Section~\ref{system} describes the overview and setup of \ac{LAL} system. Section~\ref{sec:black_assessment} provides rule-based, and LSTM-based prediction approaches to quantifiably evaluate the feasibility and performance of our system. We discuss the application of \ac{LAL} in 5G and NextG software stacks in Section~\ref{conclusion}.

\section{System Design}\label{system}
\subsection{System Overview}
\begin{figure}[!t]
\centering
\includegraphics[width=0.5\textwidth]{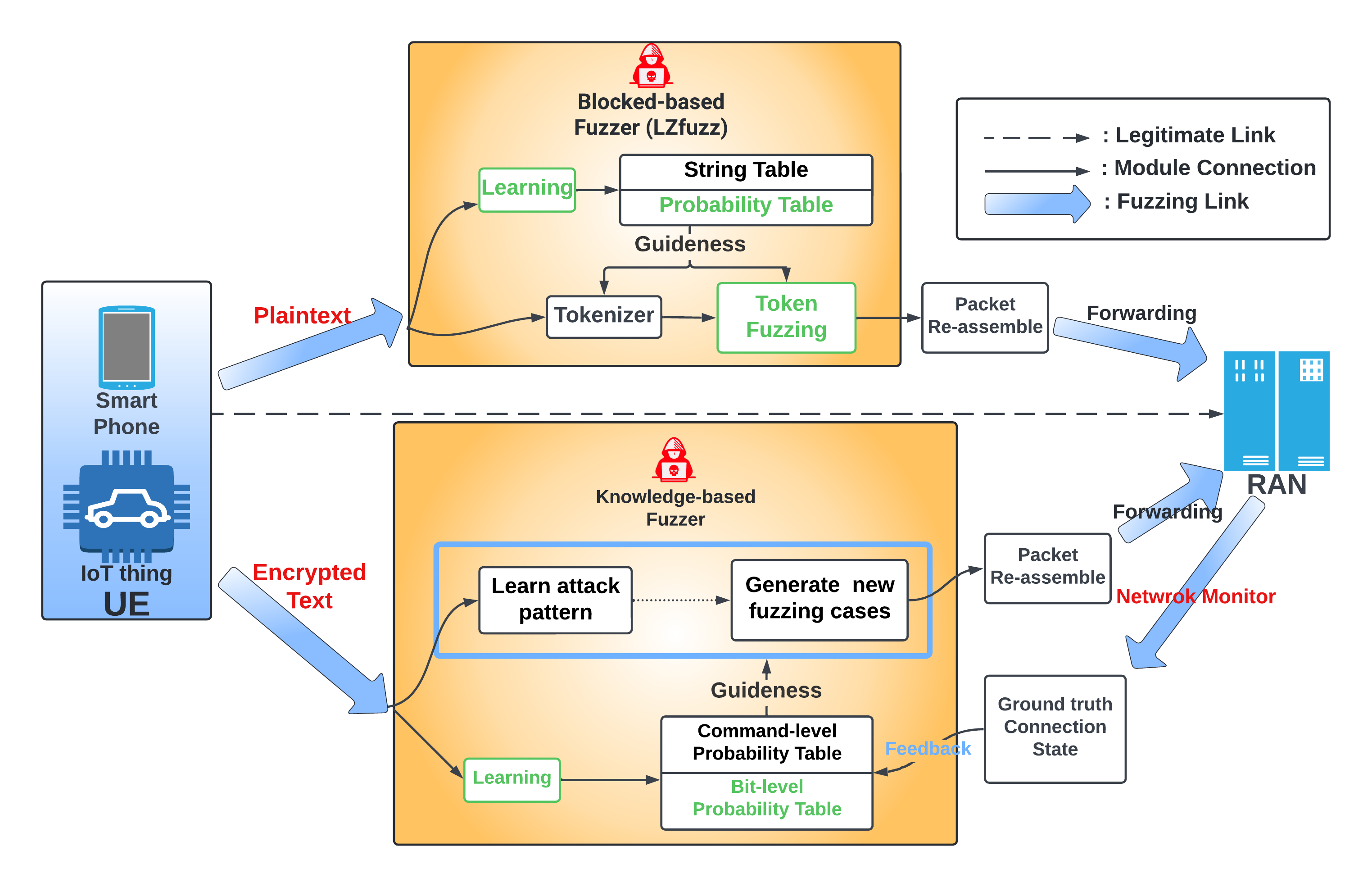}
    \caption{Overview of 5G fuzz testing methods.}
 \label{fig:system_architecture}
\end{figure}

The proposed system is scenario-adaptive to different levels of knowledge background, from no knowledge (black-box) to thorough knowledge (white-box) about protocols and Fig.~\ref{fig:system_architecture} shows the architecture of the proposed scenario-adaptive fuzzing system. First, attack model configuration is required as input, where we can define the security goals and target high-risk protocols or modules in a specific software stack based on contextual information and domain knowledge. Then, Given the input, the system could identify fuzzing locations and generate appropriate attack models. For example, the model would take the black-box \ac{LAL} strategy when the attack configuration is no knowledge, and the white-box \ac{SoAL} strategy would be selected as the attack model when the attack configuration is thorough knowledge. Finally, based on the attack model, the fuzzing strategy function will generate the fuzzing sequences ordered by the priorities. The output of the system contains the identification of high-risk states and transactions, the detected vulnerabilities, and the prediction of the vulnerable path. The consistency of our detected vulnerabilities with existing exploits, which will be illustrated in the following sections, proves the feasibility and efficiency of our system.

Based on the format and impact, we focus on the validity and legitimacy of commands. Legitimacy indicates whether the command will pass protocol and cryptographic checkers, and validity represents whether the command will lead to a threat. Correspondingly, commands mainly fall into three classes: valid states, illegal or invalid states, and other logical states, the relationship of which is shown in Fig.~\ref{fig:states}. Most command-level fuzzing states can be regarded as legal states because all command-level states are collected from regular connections. However, the validity of command-level fuzzing states can only be decided based on the result of connections, in which valid means fuzzing state has no influence on protocol stack and invalid means fuzzing state will lead to a threat or vulnerability. On the contrary, bit-level fuzzing states contain both illegal and legal states. Therefore, we use the source code interpreter as the integrity checker to identify whether the bit-level fuzzing states are legal or illegal. As for the validity of bit-level fuzzing states, we take the same measurement approach with command-level fuzzing states to label.

In this paper, we focus on neither illegal nor valid states (represented as the yellow zone in Fig.{~\ref{fig:states}}) in the RRC protocol. We disregard easily identifiable valid and illegal states, typically marked as successful and rejected connections, respectively. The states neither illegal nor valid, in which we are interested, are defined as legal and invalid. These states cannot be detected by the system through the designed exception-handling mechanism and entail unexpected vulnerabilities. In particular, we discovered 129 vulnerabilities that are actual implementation flaws in the RRC protocol. In practice, an attacker can record and replay commands exchanged between UE and the base station to establish a fake UE's connection with the base station, thus enabling a Man-in-the-Middle attack. Furthermore, by changing bits over the air, the attacker can launch a DoS attack by hijacking legitimate users.

\subsection{Hardware and Software Setup}
\begin{figure*}[!t]
\centering
\includegraphics[width=\textwidth]{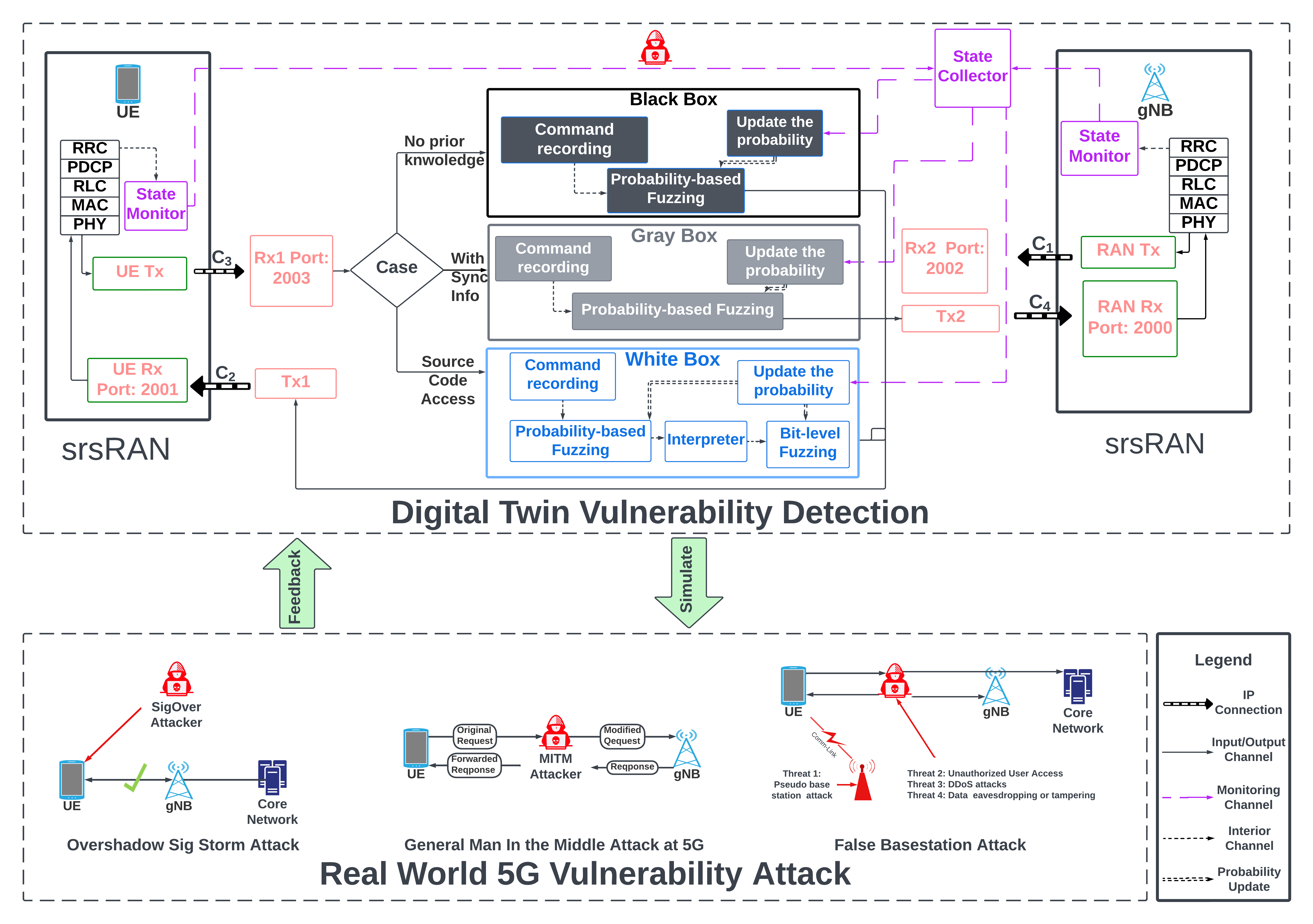}
    \caption{Digital engineering view for 5G vulnerability and unintended emergent behavior detection.}
    \label{fig:system_conf}
\end{figure*}

Fig.~\ref{fig:system_conf} shows the designed srsRAN-based \ac{MITM} digital twin model to simulate emergent behavior and wildly unexpected emergent behaviors usually occurring during legitimate physical communications. In our digital twin model, fundamental functions of \ac{UE} and \ac{gNB} are implemented by srsRAN. Furthermore, we use \ac{ZMQ}, an asynchronous socket message-transfer framework implemented with TCP protocol, as the substitute for wireless communications between \ac{UE} and \ac{gNB} in the digital twin. Then we set up a \ac{MITM} relay, which can listen and forward the socket messages between \ac{UE} and \ac{gNB}, to represent attackers in our proposed digital twin model. As for \ac{CN}, we use Open5GS to achieve all necessary functions of 5G protocols.

As the core of the proposed digital twin model, \ac{MITM} relay is responsible for message listening, modification, and recording. The following is the detailed structure of our proposed \ac{MITM} relay implementation:
\begin{itemize}
    \item \textbf{Message listening.} In Up\_link channel, the proposed \ac{MITM} relay can listen to messages from \ac{UE} by port 2003 following TCP protocol and forward the \ac{UE} messages to \ac{gNB} by port 2000. Same in the downlink channel, the proposed \ac{MITM} relay can listen to port 2002 to get messages from \ac{gNB} and forward them to port 2001 of \ac{UE}.
    \item \textbf{Message modification.} Based on the fuzzing probability system, illustrated in the following section, the relay will take command-level and bit-level fuzzing strategies to modify the message to detect vulnerabilities.
    \item \textbf{Message recording.} We build a database, shown in Fig.~\ref{fig:er}, to store the history of messages, which we listen from \ac{UE} and \ac{gNB} chronologically. And our database will also record fuzzed cases and the status of each connection attempt.
\end{itemize}
With the updated probability system by status monitors in \ac{UE} and \ac{gNB}, the relay can efficiently learn the threat patterns and detect the vulnerabilities. Not limited to \ac{MITM} attacks, the relay can also simulate overshadowing attacks or false \ac{BS} attacks by message modification. Our proposed relay can simulate any physical wireless attacks, which proves our relay is qualified for the digital twin of attackers in the real world.

\begin{figure}
    \centering
    \includegraphics[width=0.5\textwidth]{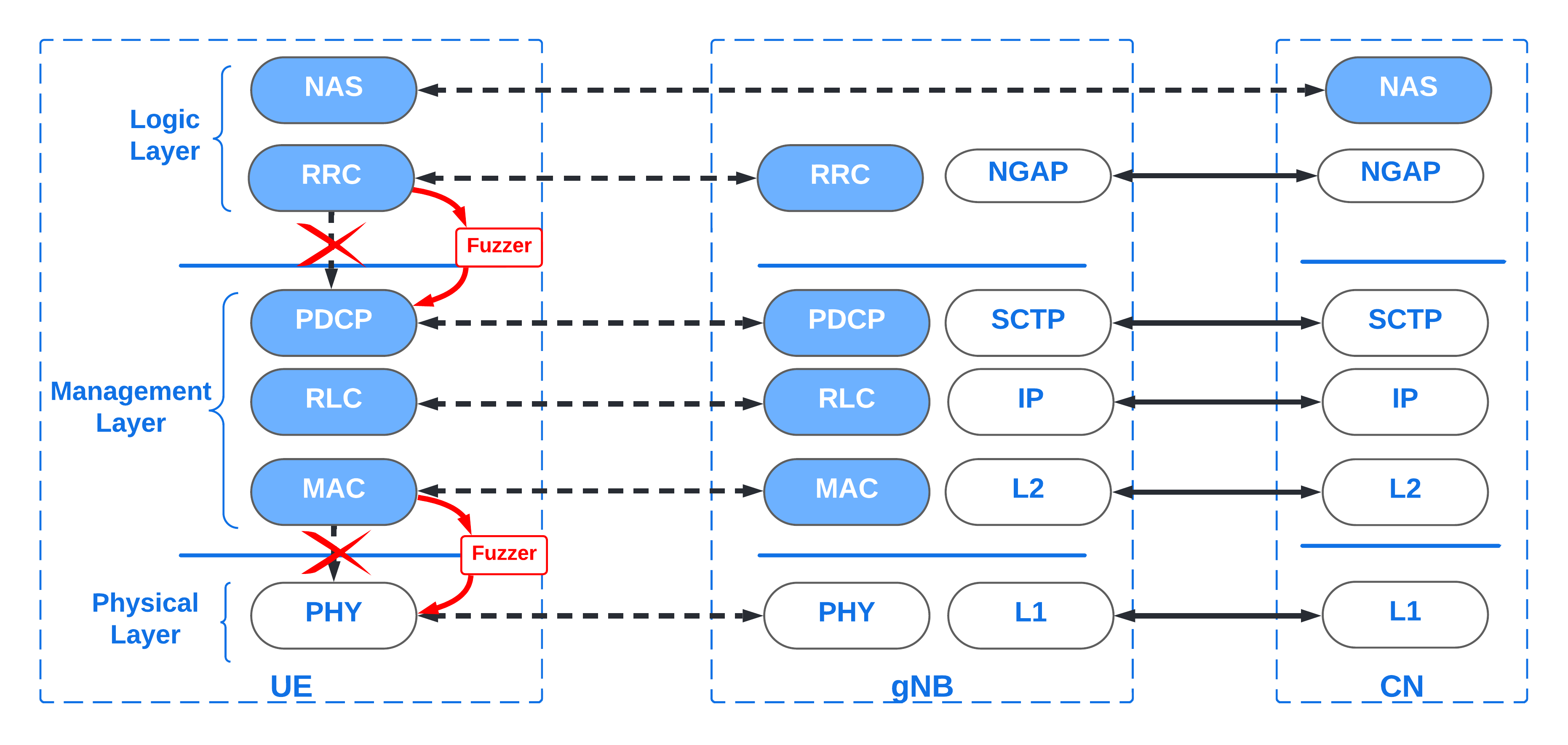}
    \caption{Fuzzing cross different layers.}
    \label{fig:fuzzing_layer}
    \vspace{-15pt}
\end{figure}

As shown in Fig~{\ref{fig:fuzzing_layer}}, we apply the relay model across different layers. In particular, we apply our fuzzing strategies to the RRC layer and the MAC layer. The RRC layer represents the logic layer connected to the management layer, while the MAC layer represents the management layer connected to the physical layer. We conducted fuzzing in the RRC layer on a simulation platform, where all identifiers can be modified at the function level. However, fuzzing in the MAC layer was performed on an over-the-air platform to directly change commands at the bit level. By comparing results from these layers, we provide a more detailed mechanism analysis of the attack model based on the inherent nature of different layers. For instance, the RRC layer regenerates the checksum for integrity checks after command-level or bit-level fuzzing, whereas the MAC layer directly forwards the fuzzed message without generating the corresponding new checksum.

The virtual relay mode in our platform enables the detection of vulnerabilities through abstract and agile methodologies. It's also vital to evaluate real-world performance and potential vulnerabilities for a thorough assessment. Our platform's realistic testing environment allows for detailed analysis of vulnerabilities, their impacts, and the system's resilience to various attack scenarios. In the OTA mode, depicted in Figure 5, we use srsRAN configured as the UE to control the USRP B210 device, facilitating communication with the Amarisoft Call Box, which acts as both the gNB and CN components. The Amarisoft Call Box can be substituted with srsRAN for an open, programmable Radio Access Network (RAN) and Core Network. However, current challenges with the stability and lack of performance benchmarks in open-source software-defined radio platforms like srsRAN, which hinder the practical application of research in industry, lead us to use the Amari Call Box as our benchmark framework. We retrieve communication logs from the Amarisoft Call Box via SSH through a switch, ensuring no impact on connection quality and performance during tests. This setup ensures seamless integration of components, efficient data exchange and analysis, and provides a reliable, comprehensive environment for our experimental investigations.

\begin{figure}
    \centering
    \includegraphics[width=0.5\textwidth]{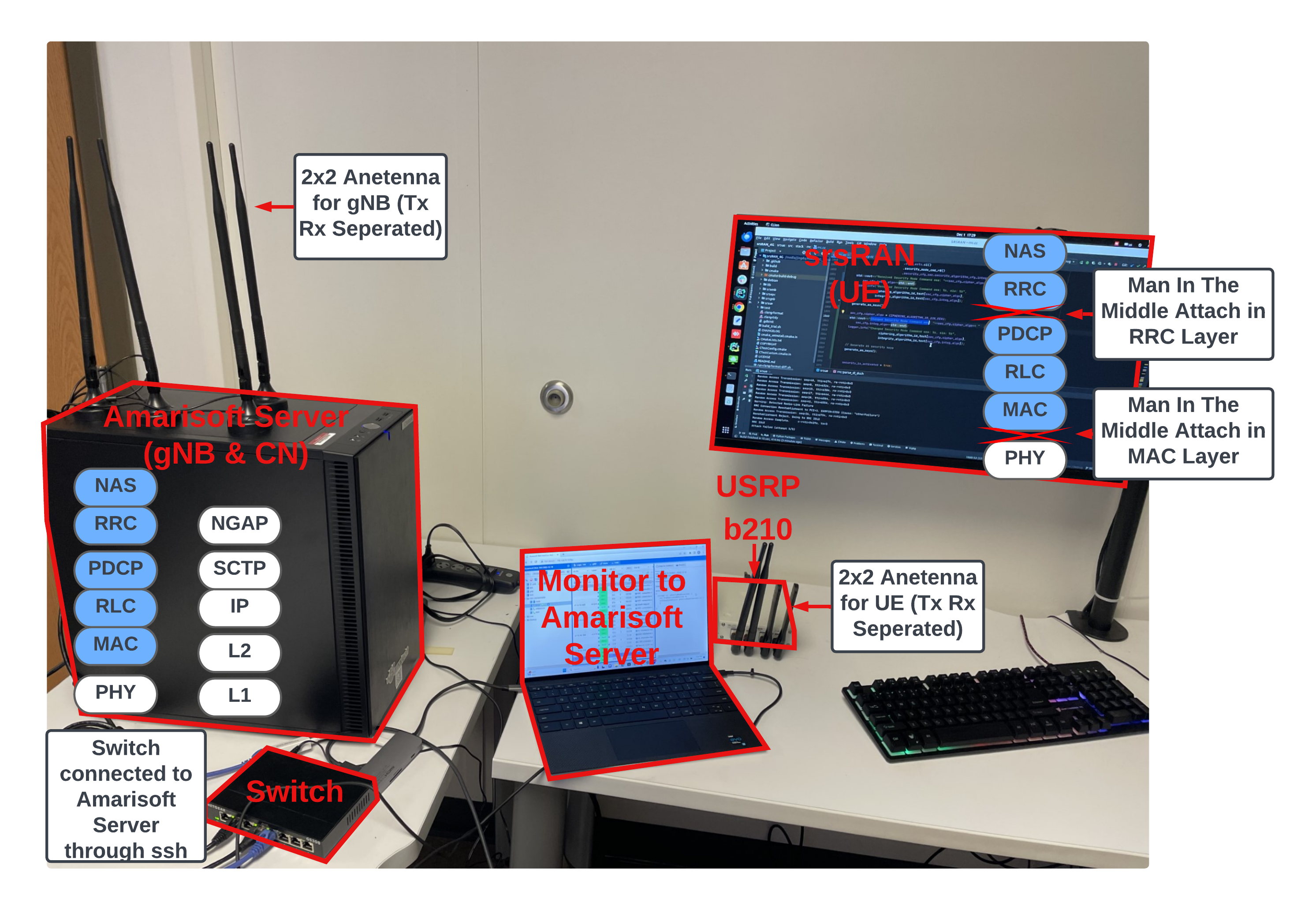}
    \caption{ Over the Air (OTA) Mode Experimental Setup and Configuration}
    \label{fig:devices}
\end{figure}

\section{\ac{LAL} Command-level Strategy}
When our proposed system has no prior knowledge or understanding of protocols, the system will try to detect and predict vulnerabilities without any domain knowledge. All commands received by our relay will be encrypted and Fourier transformed. Even under such conditions, the system proves the ability to detect vulnerabilities in black-box environment. Especially to discover and mitigate vulnerabilities and unintended emergent behaviors in the 5G stack with sufficient automation and adequate scalability, we design a protocol-independent Listen-and-Learn (\ac{LAL}) based fuzzing system.

The proposed \ac{LAL} fuzzing system is designed to target the \ac{RRC} protocols, which are recognized as one of the most critical cellular protocols for radio resources management \cite{Potnuru2021Berserker:5G}. The \ac{RRC} protocol configures the user and control planes according to the network status and allows the implementation of radio resource management. We fuzz the \ac{RRC} protocols with the srsRAN 5G platform and the tunneled \ac{NAS} protocols through message reorder, injection, and modification. We implement two-dimensional fuzzing--- command-level and bit-level, but we focus on command-level fuzzing in this work. 
We identify high-risk attack paths by generating the state-Transaction graph from the command-level fuzzing results. We further perform timely high-risk scenarios prediction with state Transaction based \ac{LSTM}.

By embedding the message exchange sniffer and the \ac{LSTM} based prediction model in virtual wireless simulation, our \ac{LAL} fuzzing system is capable of automatically and efficiently determining the command-level fuzzing message according to the states of \ac{UE} and \ac{gNB}. 
Besides, as the designed framework is protocol independent, it can be quickly adapted and transferred to new-released code stacks and protocols. For example, it can be easily turned into a hybrid design to provide provable assurance and formal threat detection for 5G software stacks by combining deterministic detection approaches such as formal methods. Finally, we incorporate an LSTM model based on rapid vulnerability detection. This prediction model enables proactive defenses against potential attacks through learning the early-stage abnormal state transaction paths. In short, the designed \ac{LAL} fuzzing system can be applied to 5G and NextG architectures (e.g., ORAN) for real-time vulnerability and unintended behavior monitoring, prediction, detection, and tracking. 

\subsection{State Recording}

The authentication and authorization scheme in 5G, being viewed as a finite state transaction, enables the graphics-based analysis to identify the pattern of the risks. During the fuzzing, the recorded states include the following information: `message time', `original bytes', `RRC channel', `message type', and `physical channel'. `Message time' represents message sending time, and `RRC channel' indicates which protocol we should use for message decoding. The \ac{RRC} procedure of the \ac{UE} can be uniquely identified with the 'Message type' \cite{TSGR2018TS15}. Due to a lack of domain knowledge about encryption and transformation, we use a general 5G protocol, pycrate \cite{pycrate}, to interpret the first six hex values and select the cross mapping of interpreted `message type' as the identifier to a state. Even though this interpretation approach can not provide the correct translation of command, the feasibility of vulnerability detection and prediction can still be proved in Sec \ref{sec:black_assessment}. Besides, the `rrcConnectionSetupComplete' message is used as the identifier of the successful completion of \ac{RRC} establishment. When the monitor in gNB detects the `rrcConnectionSetupComplete' message, the testing case will be terminated and labeled as a successful connection. When the monitor in gNB cannot detect the `rrcConnectionSetupComplete' message within a predefined timeout limit ($600$ seconds in the proof-of-concept experiments), it is considered a failed connection.

\begin{figure}[!t]
\centering
    \includegraphics[width=0.5\textwidth]{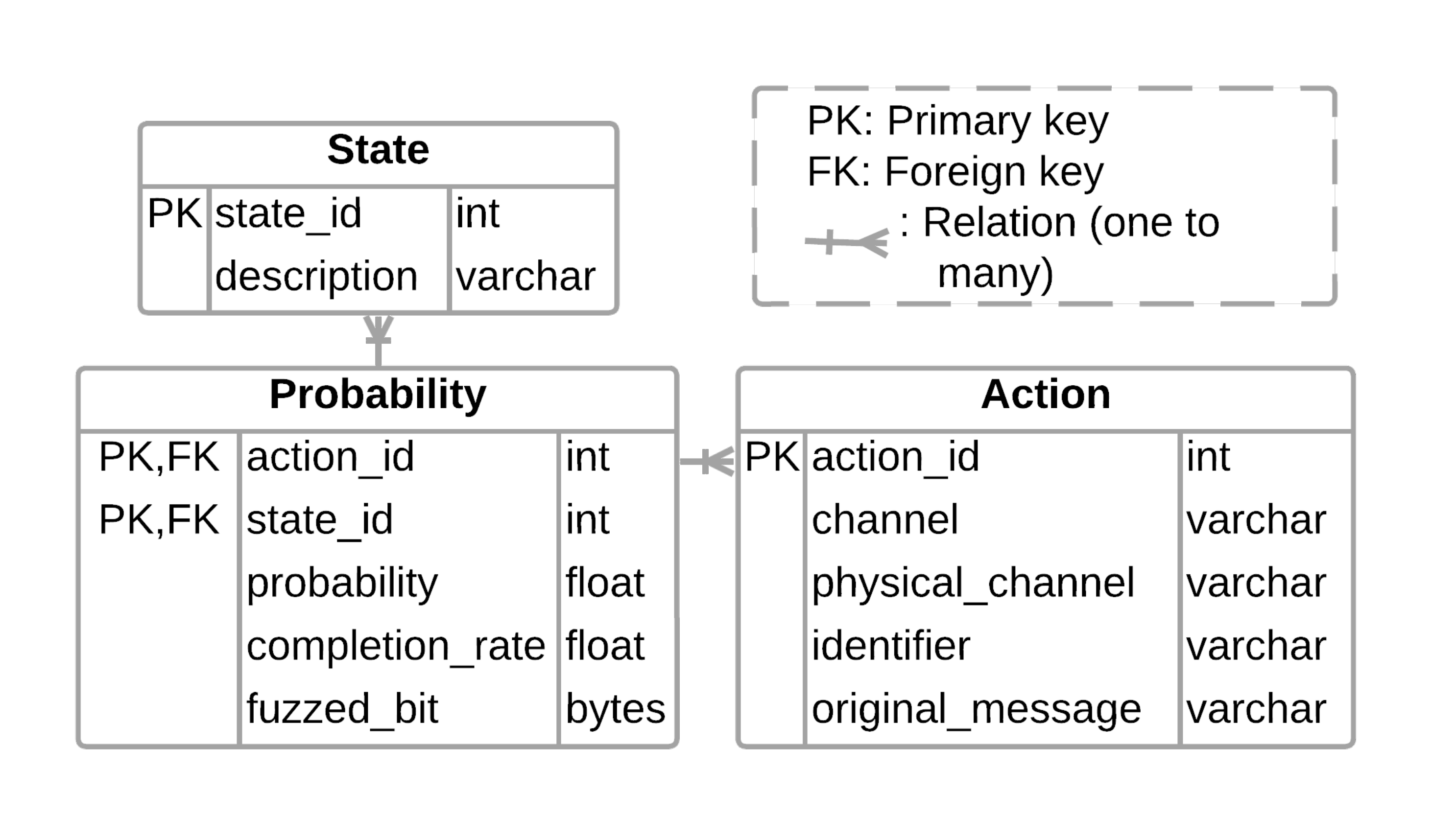}
    \caption{Structure of the database}
    \label{fig:er}
\end{figure}

We build a database to record the state and fuzzing cases. Fig.~\ref{fig:er} illustrates the structure of the database. The foreign keys in Fig.~\ref{fig:er} are generated from the primary keys of the refereed table like `action\_id' is generated from the primary key `action\_id' in table Action. The statements of each table are described in the following:
\begin{itemize}
    \item \textbf{State.} Each state represents the state of RRC status. For each message sent from either UE or eNB, the system will update the description of sent commands in the table state.
    \item \textbf{Action.} Each action item records parsed message, whose channel and physical channel represent where the commands come from.
    \item \textbf{Probability.} Each probability item records the probability of fuzzing cases for corresponding states and actions. If one fuzzing case leads to RRC connection failure, the probability increases. The completion rate indicates the record of bit-level fuzzing and will be none if there is only command-level fuzzing.
\end{itemize}

\subsection{Command-level Fuzzing Strategy}\label{sec:fuzz_stategy}
Command-level and bit-level fuzzing are two primary protocol fuzz testing approaches. They are also common approaches for protocol attacks. Compared to bit-level attacks, which require time and frequency synchronization along with information in \ac{UE} profiles, command-level attacks are usually low-cost and less information needed. Hence, we primarily focus on command-level fuzzing to detect vulnerabilities under a black-box environment. 

For fuzzing purposes, the \ac{LAL} observes and collects the exchanged legitimate messages and saves them to the fuzzing message candidate pool. At the command level, commands are replaced by other commands in the same physical channel to test whether any communication error state occurs. More specifically, fuzz testing is implemented iteratively through each case in the pool. Within each loop, we first simulate \ac{UE} and \ac{gNB} for connection initialization. Then we decode the observed message and get two primary \ac{RRC} identifiers: `interpreted message type' and `interpreted \ac{RRC} TransactionIdentifier'. If the message in this physical channel has never been observed before, we record the message and mark it with the corresponding channel. Moreover, if there is still any unapplied message replacement, we adapt this replacement and delete it in the recording. Due to the change of temporary identifiers, such as rnti, most of the replaced messages are illegal for \ac{UE} or gNB. In this way, our fuzz testing framework replaces messages with not only regular ones but also abnormal messages since the number of message permutations grows with the increasing number of cases. Due to no requirement of prior knowledge, \ac{LAL} can be quickly adapted and transferred to new-released code stacks and protocols.

                            

\subsection{Result assessment}\label{sec:black_assessment}

\subsubsection{State Based Vulnerability Prediction Model\label{sec:state}}
\begin{figure}[!t]
    \centering
    \includegraphics[width=0.5\textwidth]{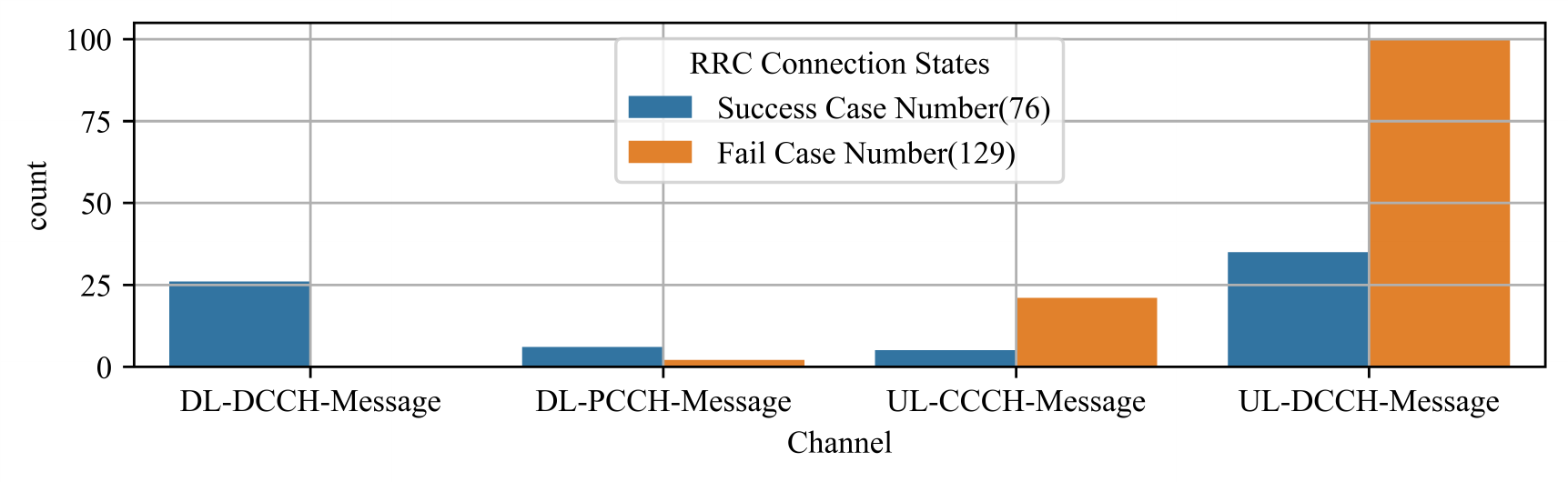}
    \caption{\ac{RRC} connection state fuzzing distribution.}
    \label{fig:distribution}
    \vspace{-15pt}
\end{figure}

As commands are listened to and selectively added to the candidate pool during the experiments, the system needs to traverse the commands in the pool with priority orders to perform command-level fuzzing. As an initial step to command priority exploration, we show the \ac{RRC} connection state distribution of commands fuzzing at various channels in Fig.~\ref{fig:distribution}, where DL and UP in the x-axis represent downlink and uplink channels, respectively. Among a total of $205$ collected fuzz testing cases with \ac{RRC} procedures, there are $76$ successful connections and $129$ failed connections. The majority of the failed \ac{RRC} connections are through uplink fuzzing. The failures in the downlink fuzzing channel are primarily caused by PCCH Messages, which are used for paging the \ac{UE}s whose cell location is not known to the network.

\begin{table}
  \caption{Fuzzing Test and State Distribution in Different Types}
  \label{Tab:amount}
  \centering
 \begin{tabular}{|c|c|c|}
  \toprule
  \textbf{Amount}&
  \textbf{Successful Connection}&
  \textbf{Failed Connection}\\
  \midrule
  \tabincell{c}{Cases\\States}&
  \tabincell{c}{76\\39}&
  \tabincell{c}{129\\7}\\
  
  \bottomrule
  \end{tabular}
\end{table}

From the distribution, we can conclude that the downlink channel protocols are more robust with unintended messages, and uplink channels are more vulnerable than the downlink. However, due to PCCH-Messages broadcast and explicit content nature among the downlink channel, its vulnerability could affect a more extensive range of communications and potentially cause a \ac{DoS} attack for all \ac{UE}s within the cell of the BS.

For the $129$ failed connection in Fig.~\ref{fig:distribution}, we identify high-risk states as those with high frequencies in failed connections. Our results show that no state only appears in failed connections but not in successful ones. We identify $7$ high-risk states with higher frequencies than the average. However, the \ac{RRC} connection failure cannot be fully covered with those high-risk states solely as rule-based detection. Therefore, we introduce transaction-based detection using the sequenced states to enhance vulnerability identification. 

\subsubsection{Vulnerability Identification via State Transaction\label{rule}}
With the sequence of fuzzing tests being executed, the system automatically generates a state transaction probability map. The probability map predicts the connection risks, and further rerouting strategies can be developed to avoid certain states and transactions that may potentially lead to \ac{RRC} connection failures. The RRC state changes from one to another are defined and recorded as a transaction. We can graphically represent the state and transaction during the \ac{RRC} procedures as the vertex and edge to ease further graphics-based analysis for risk identification and prediction. 

The occurrence of the state-transaction cases can be used for the rule-based prediction of failed connections. Fig.~\ref{fig:frequency} shows the state transaction frequency on successful and failed connections, from which we can observe $7$ high-risk state-transaction cases that almost only occur in failed connections. This rule-based prediction using transaction frequency is more trustworthy compared to the prediction based on state frequency. Because among all $7$ high-risk state transactions, there is only one transition: from state 0 to state 2, occurring in a total of 76 successful connections for only one time. By further looking into the results of Fig.~\ref{fig:frequency}, we observe that $70.54\%$ failed connections include at least one high-risk transaction. Hence, given that recall equals $70.54\%$, a more accurate algorithm is necessary to identify and predict \ac{RRC} connection failures. In the following, we presented an \ac{LSTM} based vulnerability prediction on providing high-confidence predictions.

\begin{figure}[!t]
     \centering
    \begin{tabular}{ c}
       \includegraphics[width=0.4\textwidth]{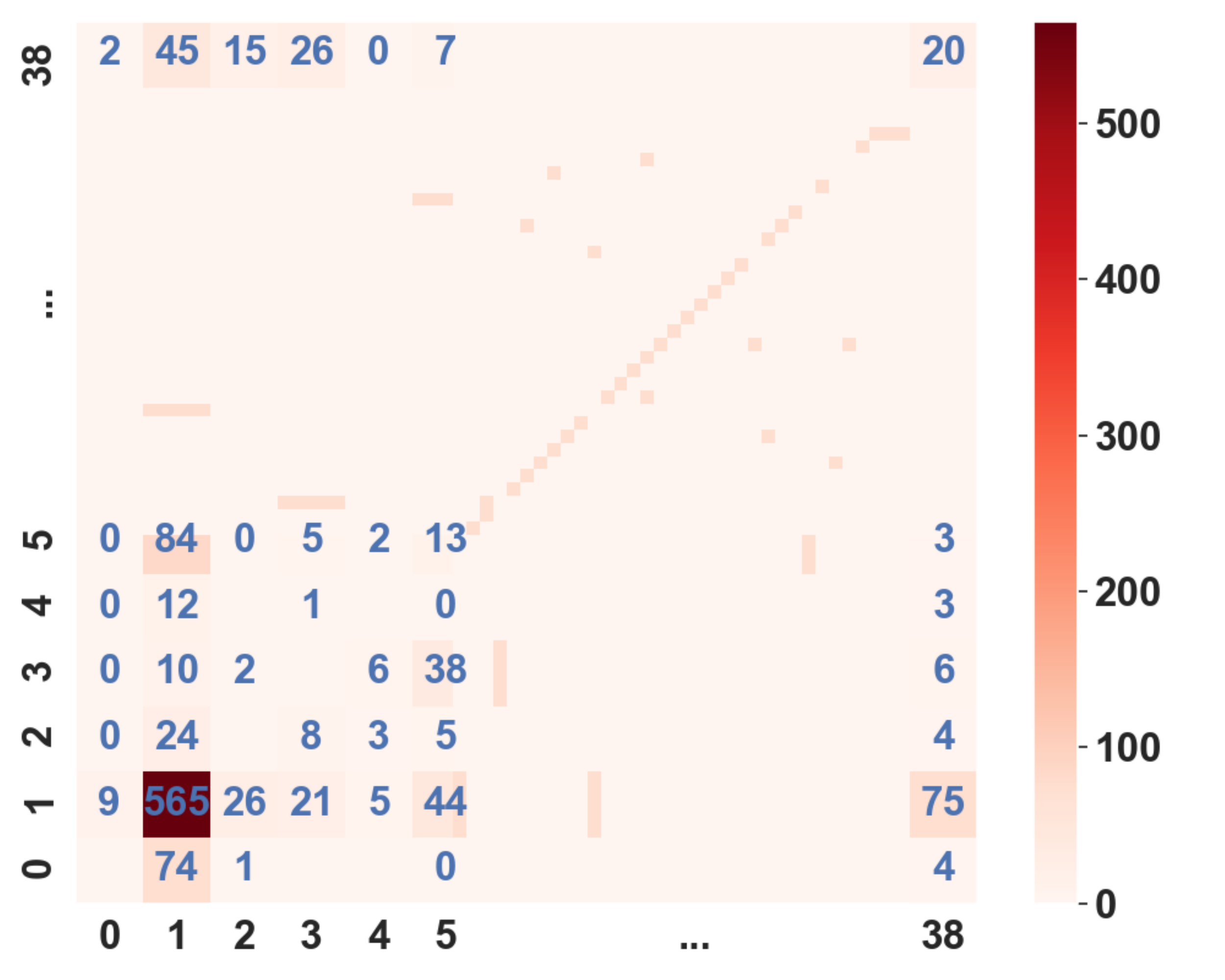}  \\\small (a)\\
        \includegraphics[width=0.4\textwidth]{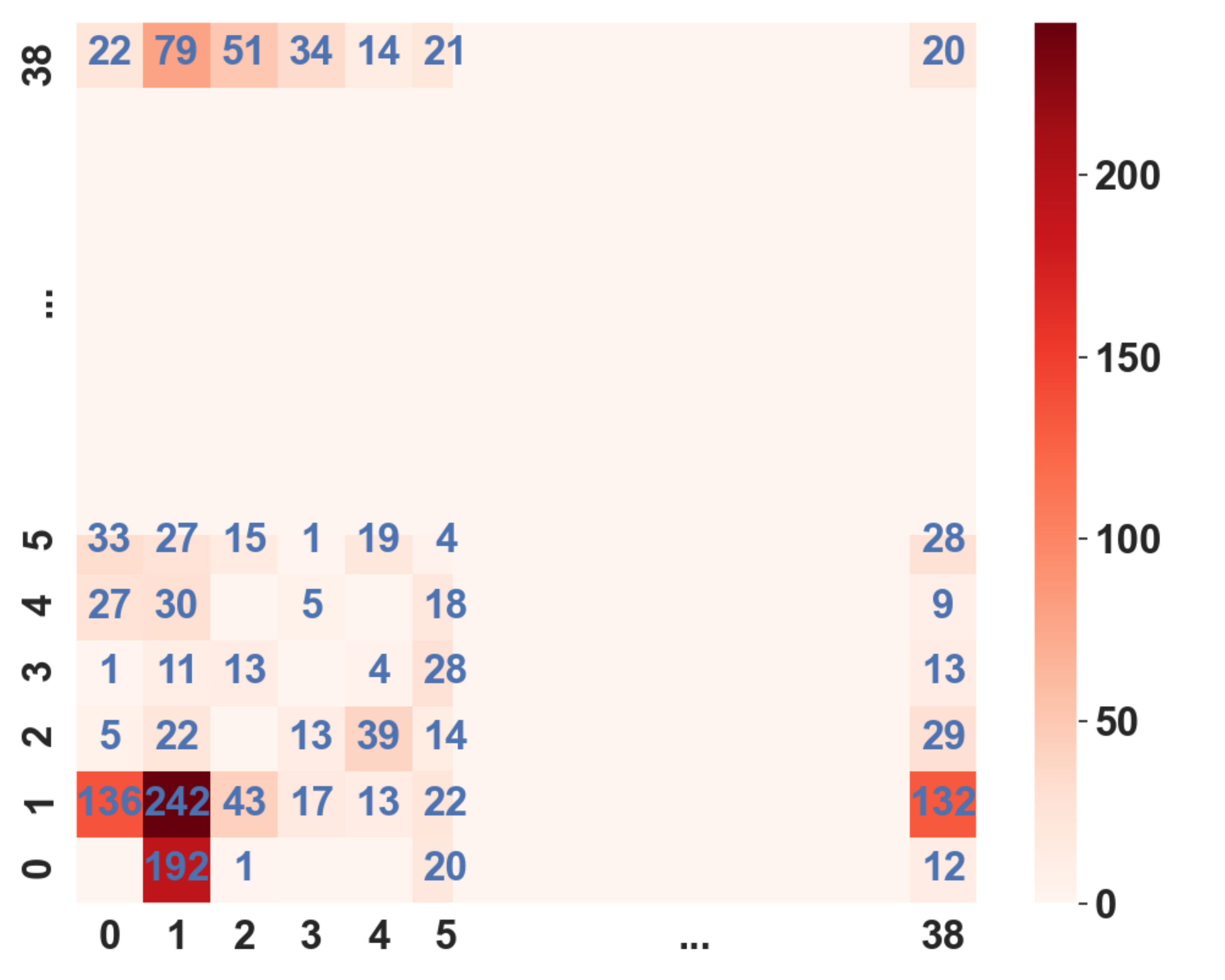}\\
      \small (b)\\
     
    \end{tabular}
    \caption{State transaction frequency on: (a) successful connection; (b) failed connection.}
    \label{fig:frequency}
    \vspace{-15pt}
\end{figure}

\subsubsection{\ac{LSTM} Based Vulnerability Prediction }

The results from Sec. \ref{sec:state} and  \ref{rule} show that the statistic and rule-based classification can only achieve recall up to $70.54\%$, which is unreliable in practice. Therefore, we design an \ac{LSTM} based vulnerability prediction model for reliability enhancement and early prediction. With the early prediction of \ac{RRC} connection failures, we can enable \ac{RRC} state rerouting strategy to avoid the failures. 

We define the input of \ac{LSTM} as the sequenced states from the fuzzing occurrence and the length of the sequenced states as the cut-off length. The cut-off length determines how long a state transaction path can be sufficient to meet the expected accuracy. Two approaches are used to specify the cut-off length: \emph{duration from beginning} and \emph{number of states from beginning}. To avoid overfitting, we use $20\%$ of the dataset as testing data and $0.001$ as the learning rate of the model. Moreover, for each input size, we average the accuracy, precision, and recall over $100$ runs. Each run includes $30$ epochs, and each epoch includes $10$ batches. As shown in the performance evaluation of both approaches in Fig.~\ref{fig:frequency}, accuracy grows with the increasing number of steps or times and increases sharply after the $8$-th step.

\begin{figure}[!t]
     \centering
    \begin{tabular}{ c}
       \includegraphics[width=0.45\textwidth]{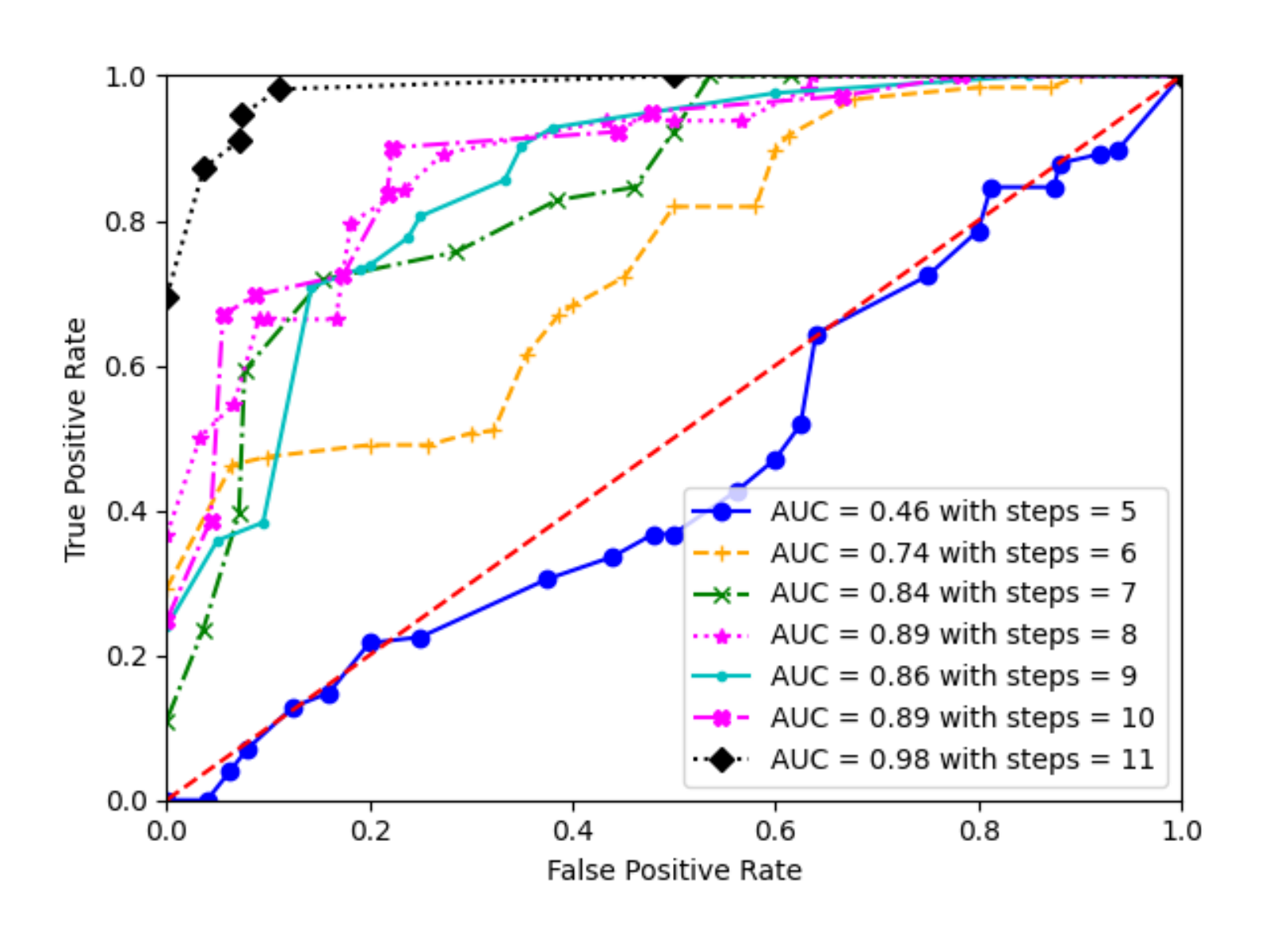} \label{fig:roc_steps} \\\small (a)\\
        \includegraphics[width=0.45\textwidth]{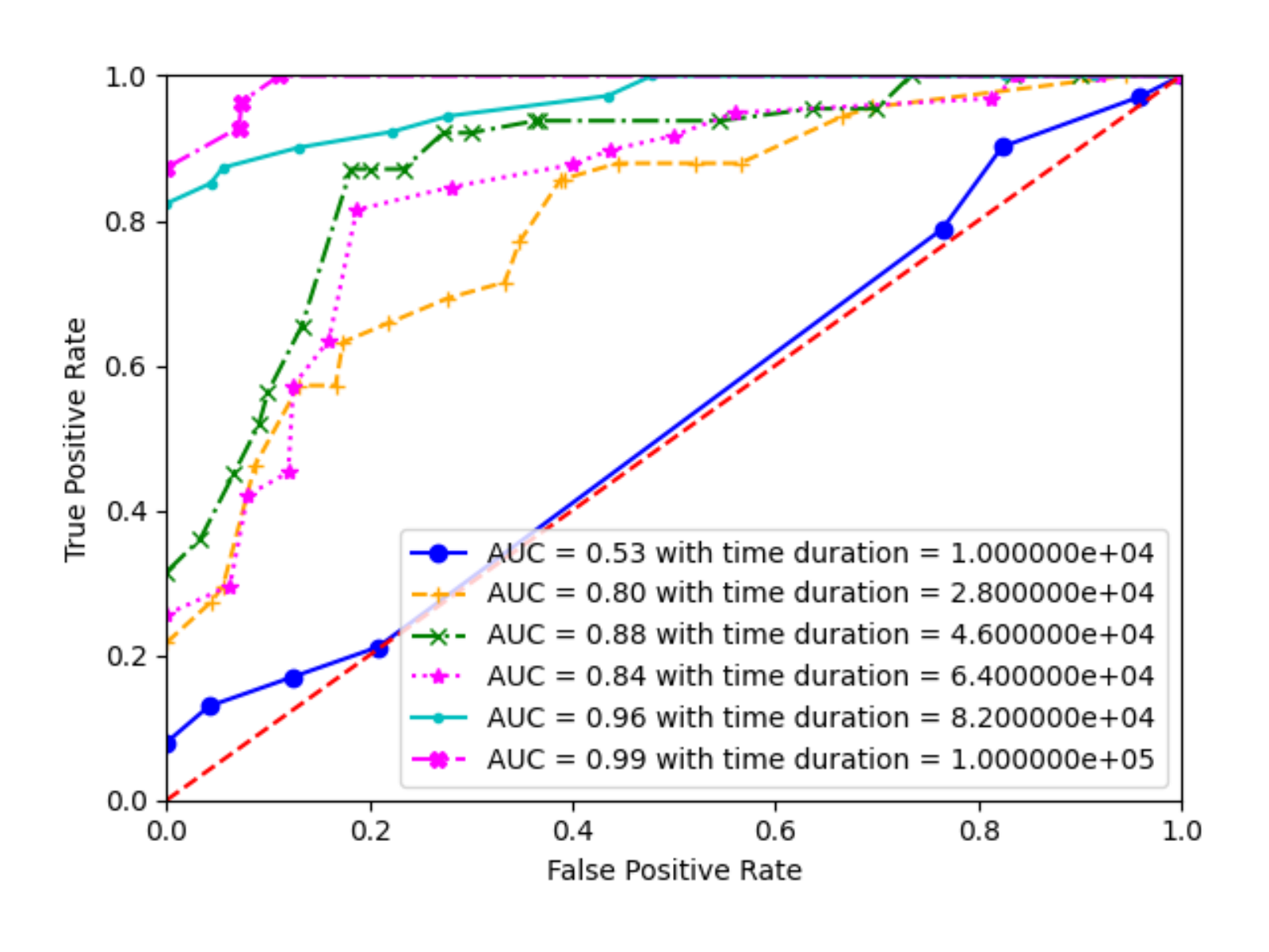}\label{fig:roc_times}\\
      \small (b)\\
     
    \end{tabular}
    \caption{Receiver Operating Characteristics (ROC) analysis of LSTM over: (a) steps; (b) duration.}
    \label{fig:frequency}
    \vspace{-15pt}
\end{figure}


To balance the performance and reaction time, we generate \ac{ROC} curves over steps and duration to find the strategy with the least cut-off length and almost $90\%$ \ac{AUC}. From Fig.~\ref{fig:frequency}(a), $10$ steps is the optimal strategy that achieves stable $89\%$ \ac{AUC}. And we can get that $0.08s$ is the optimal strategy that achieves $96\%$ \ac{AUC}. Therefore, we take $10$ steps and $0.08$s as the input to do deeper analysis on the converge performance of \ac{LSTM}. From converge performance of \ac{LSTM}, we find that \ac{LSTM} can learn the optimal parameter in 2 or 3 epochs. The fast convergence proves that our system has the ability to learn the pattern of failed connections. The average cut-off duration of $10$-th steps is $0.072$ seconds with the accuracy achieving $89\%$, which is consistent with setting the cut-off length as $0.08$ seconds through the duration cut-off approach. The accurate and timely prediction also provides sufficient time for proactive defense before RRC connection completion or failure, with an average of $3.49$ seconds.



With an average performance of \ac{LSTM} meeting the accuracy expectation, we analyze failed predictions, including False Positive and False Negative ones, for further improvements. The following patterns are summarized as misclassified cases.

\textbf{False Positive:} When there are three messages, which are interpreted as paging messages, sent by \ac{gNB} within the first ten states in connections, the model may misclassify this connection as a failed one. Because most of the failed connections also have three paging messages in the first ten states. This pattern can be addressed by a finer definition of paging messages in future work.

\textbf{False Negative:} The cases with multiple times of interpreted $\rm{active\_set\_update}$ messages in downLink channels are classified as failed connections and lead to false alarms. A high frequency of interpreted $\rm{active\_set\_update}$ messages occur more frequently in successful connections. Cross-layer or side channel information could be applied to improve the false alarm rate in future work. 

\begin{figure}[!t]
    \centering
    \includegraphics[width=0.5\textwidth]{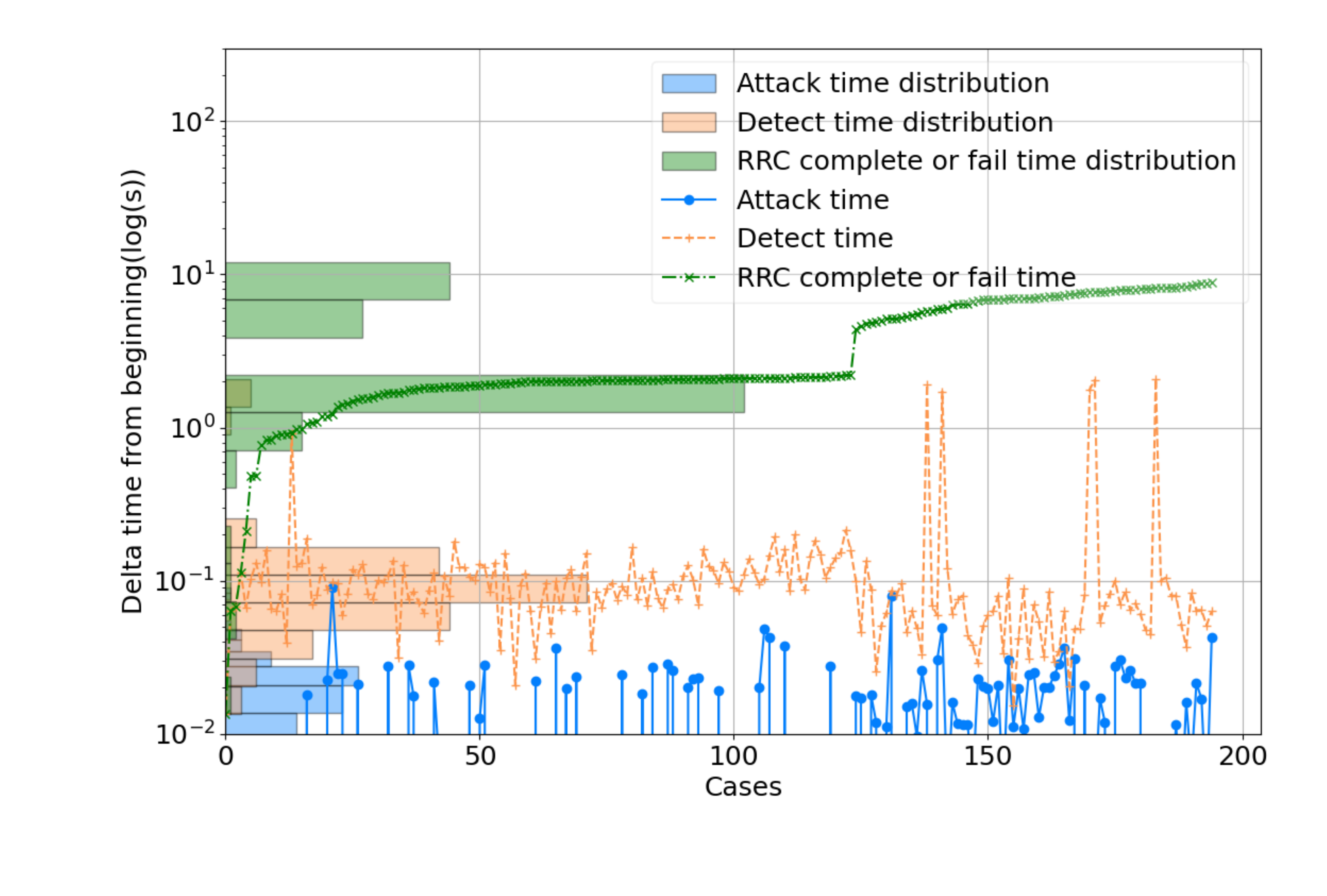}
    \vspace{-4mm}
    \caption{Comparison of detection time and completion time.}
    \label{fig:time_compare}
    \vspace{-15pt}
\end{figure}

The proposed vulnerability and unintended behavior detection system could also be applied in real-time, in compliance with \ac{O-RAN} architecture~\cite{polese2022understanding}. When deployed for real-time monitoring, timing in detection is critical to mitigate the vulnerabilities and provide assurance and high-quality communications. There are sufficient intervals, an average of $3.49$ seconds, between successful detection time and \ac{RRC} connection successful\slash failed time, as shown in Fig.~\ref{fig:time_compare}, which gives enough defense time for potential attacks. Moreover, we have tinier gaps between fuzzing occurrence time and successful detection time.

\section{\ac{SyAL} Domain Assisted Strategy}
Synchronization in the generated random identifier of each fuzzing case enables more domain information background possible for \ac{MITM} attacker, including command type and critical identifier values. Leveraging the domain knowledge, we propose a probability-based command-level fuzzing system called Synchronize-and-Learn (SyAL) to help our proposed digital twin \ac{MITM} attacker be more efficient in prioritizing and locating the more vulnerable areas. In the proof-of-concept of this study, we fix the \ac{RNTI} of \ac{UE} to keep commands synchronized in different fuzzing cases. \ac{RNTI} is the \ac{CRC} mask, which is generated in synchronization procedure and required to encode and decode \ac{DCI} message. The synchronized commands provide domain-assisted background knowledge for our digital twin \ac{MITM} attacker. Furthermore, our proposed probability-based command-level fuzzing system takes a Sync-and-Learn strategy to learn the vulnerability pattern efficiently and prioritize high-risk commend-level fuzzing cases. The result of this study proves the significance of timing in the 5G \ac{AKA}.

\subsection{Not Illegal Command-level Fuzzing in the Not-illegal Not-valid Set}
Besides the illegal command-level fuzzing in \ac{LAL}, we will continue not illegal command-level fuzzing, which contains correct identifiers and can be appropriately interpreted by \ac{UE} and \ac{gNB}, in this section. Through changing the occurrence timing of not illegal commands, we can find a path that can transfer from `green zone', valid states, to `yellow zone', not illegal and not valid states, in Fig.~\ref{fig:states}. This part of the experiment provides substantial proof of the feasibility of the listen-and-replace relay attack directed by a part of the communication context. Due to the limitation of the numerous capacity for command-level fuzzing permutation, we focus on the downlink channel, which is sent from \ac{gNB} and is more vulnerable than uplink. All messages in the downlink channel are duplicated with the same \ac{RNTI} and belong to legal or not illegal commands (the `green zone' or `yellow zone' in Fig.~\ref{fig:states}). 

\subsection{Probability-based Fuzzing Strategy}
With the domain knowledge of message types, we proposed a probability-based command-level fuzzing system to learn the vulnerability pattern efficiently and prioritize high-risk commend-level fuzzing cases. The efficiency of our proposed probability-based command-level fuzzing system outperforms traditional fuzzing systems, like brute force fuzzing.

Algorithm \ref{alg:prob_fuzz_testing} describes the detailed process of a probability-based fuzzing system. First, we build up a database to store all commands in the downlink channel, whose structure shows in Fig.~\ref{fig:er}. Then we initialize a command-level fuzzing probability matrix $D.p$ with the size of $n\times n$, $n$ is the number of commands, to represent the probability of command fuzzing cases. The value of $D.p_{i,j}$ means the probability of the fuzzing case, which changes from command $i$ to command $j$, is high-risk. After initialization, the system updates the command-level fuzzing probability matrix based on the fuzzing result after each fuzzing test. The command-level fuzzing probability matrix update follows the independent rule: the system can only update the row and column corresponding to fuzzed commands. Moreover, in each fuzzing case, the system uses the proposed digital twin \ac{MITM} attacker to generate fuzzing cases based on the value of the command-level fuzzing probability matrix $D.p$.   

\begin{algorithm}
    \begin{flushleft}
    \textbf{Input}: $I_1$, $I_2$\\
    \textbf{Output}: $O_1$, $O_2$\\
    \textbf{Initialization}: Database $D$, $ratio = 0.1$, $\alpha = 0.5$, $D(p,p^{\prime}) \gets$ probabilities for replacement commands, $fuzzed \gets False$
    \begin{algorithmic}[1]
    
    \caption{\ac{SyAL} Fuzzing Testing}
     \label{alg:prob_fuzz_testing}
        \PROCEDURE{Fuzzing\_Relay}{$I_1$}
            \WHILE{exist\_fuzzing(D)}
                \STATE start\_simulator()
                \STATE command\_history $c \gets$  []
                    \WHILE{$!\, time\_out()$}
                        \IF{$a$ = received\_action($I_2$)}
                            \IF{$a$ = rrc\_Connection\_Complete}
                                \STATE break
                            \ENDIF
                            \STATE update $D$($a$)
                            \IF{fuzzed}
                                \STATE continue
                            \ELSE
                            \STATE $a^{\prime}$ = $random(D.p[a,:])$  \Comment{Weighted random}
                            \STATE $send\_as\_O_2$($a^{\prime}$)
                            \STATE $c$ += ($a$,$a^{\prime}$)
                            \STATE $D.p^{\prime}.update.$($a^{\prime}$)
                            $fuzzed$ = $true$
                            \ENDIF
                        \ENDIF
                    \ENDWHILE
                \IF{connection\_failed()}
                
                        \STATE $D.p$[$a$,$a^{\prime}$] += $D.p[a,a^{\prime}] \times \alpha$ 
                \ELSE
                        \STATE $D.p$[$a$,$a'$] -= $D.p[a,a'] \times \alpha  \times ratio$
                \ENDIF  \Comment{Update database as $O_2$}
            \ENDWHILE
        \ENDPROCEDURE
    \end{algorithmic}   
    \end{flushleft}
   
\end{algorithm}

\subsection{Result Assessment}

\begin{figure}[!t]
     \centering
    \begin{tabular}{ c}
       \includegraphics[width=0.45\textwidth]{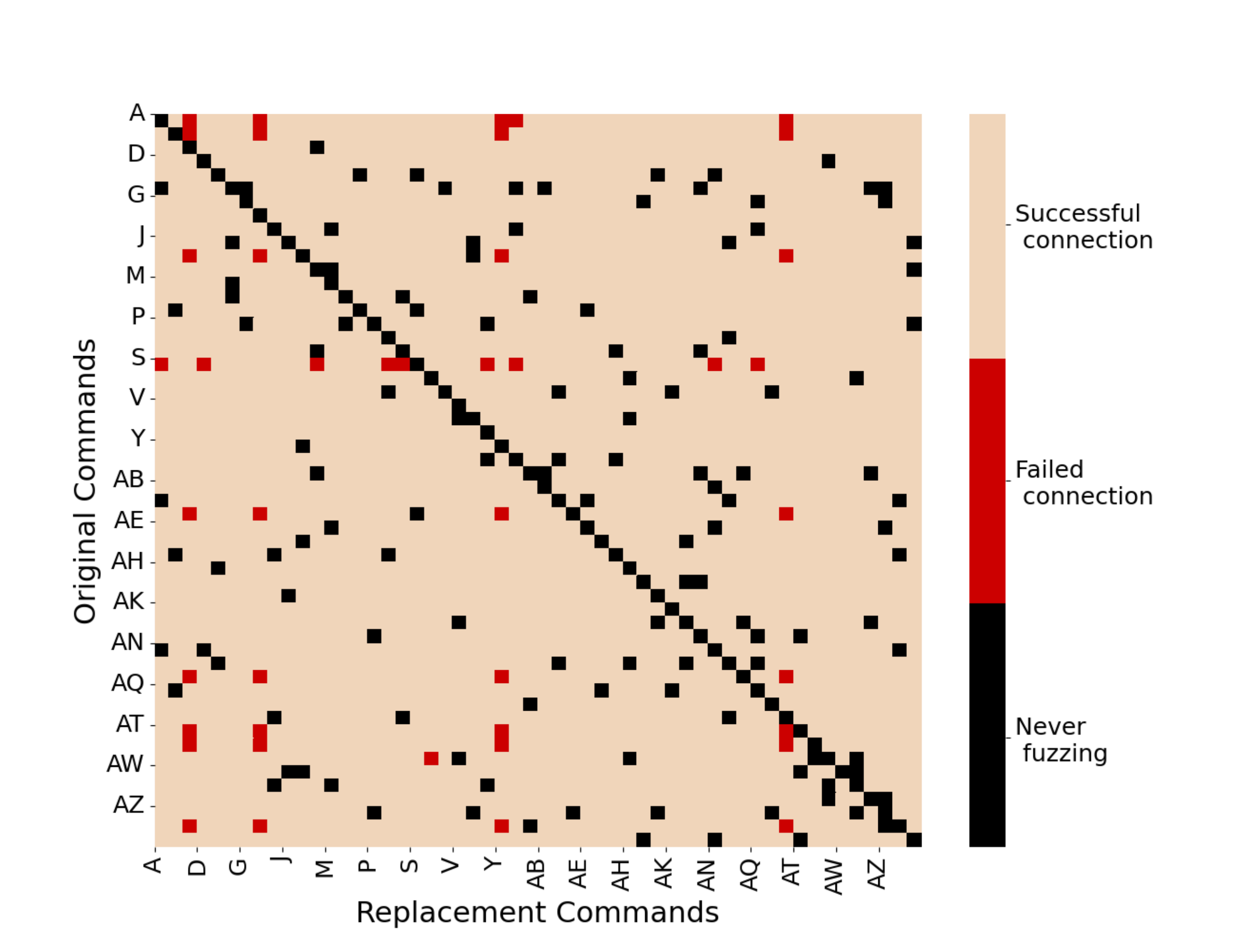} \label{fig:roc_steps} \\\small (a)\\
        \includegraphics[width=0.45\textwidth]{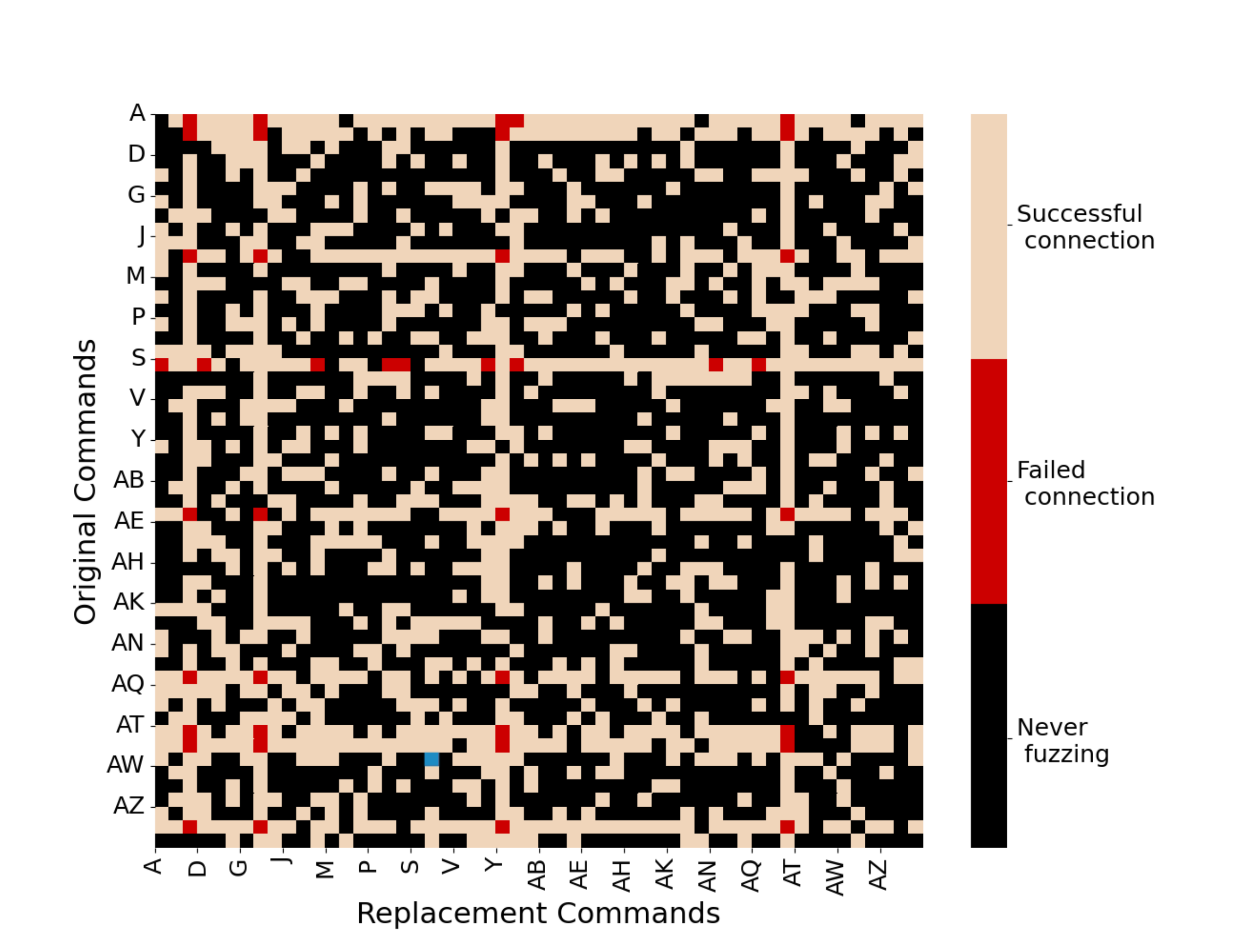}\label{fig:roc_times}\\
      \small (b)\\
     
    \end{tabular}
    \caption{Fuzz testing tracks on downlink channels of: (a) random fuzzing; (b) probability-based fuzzing.}
    \label{fig:fuzzing}
    \vspace{-15pt}
\end{figure}

\begin{table}
  \caption{Legal Command-level Fuzzing Test Distribution in Different Types}
  \label{Tab:legal_amount}
  \centering
 \begin{tabular}{|c|c|c|}
  \toprule
  \textbf{Amount}&
  \textbf{Successful Connection}&
  \textbf{Failed Connection}\\
  \midrule
  \tabincell{c}{Cases\\States}&
  \tabincell{c}{3037\\43}&
  \tabincell{c}{54\\13 (high-risk)}\\
  
  \bottomrule
  \end{tabular}
\end{table}
As mentioned in Table \ref{Tab:legal_amount}, there are 3080 possible fuzzing cases and only 43 vulnerabilities.
In Fig.~\ref{fig:fuzzing}, we present fuzz testing tracks of two fuzzing strategies: random fuzzing and probability-based fuzzing, until all vulnerabilities are found. In Fig.~\ref{fig:fuzzing}(a), there are 2811 fuzzing cases fuzzed by random fuzzing strategy until all vulnerabilities founded. However, the probability-based fuzzing strategy takes only 1027 fuzzing cases to get all vulnerabilities in Fig.~\ref{fig:fuzzing}(b). Therefore, we can easily conclude that the probability-based fuzzing strategy can locate the vulnerabilities much more efficiently than the random fuzzing strategy.

\begin{figure}[!t]
    \centering
    \includegraphics[width=0.5\textwidth]{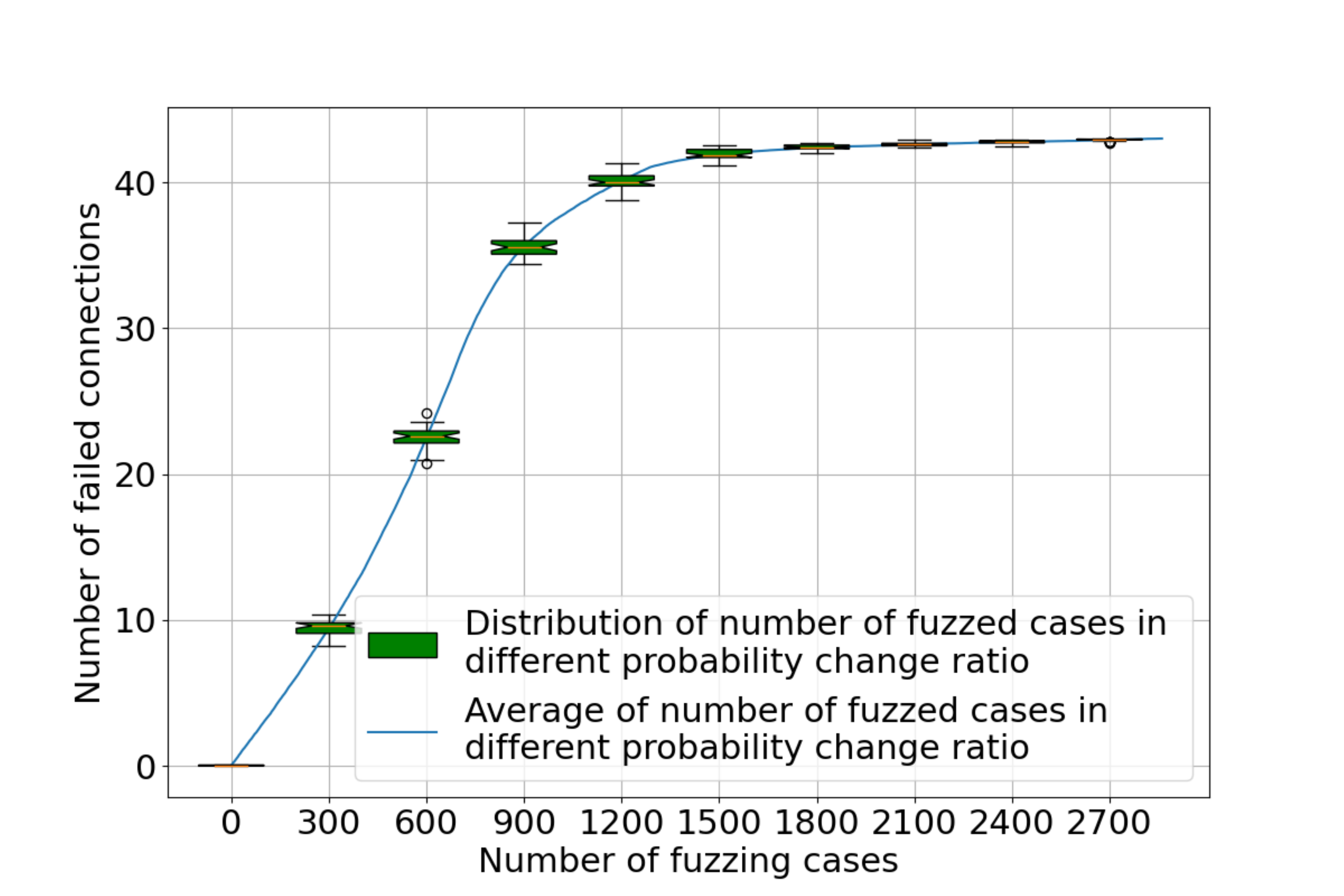}
    \caption{Sensitivity analysis of different probability ratio.}
    \label{fig:sentivity}
    \vspace{-15pt}
\end{figure}

Further, we use the hyper-parameter strategy with the permutation of change percentage $\alpha$ from 0.1 to 2 and failed attenuation $ratio$ from 0.9 to 0.1 to get the optimal parameters set for probability-based fuzzing strategy. However, the result of the hyper-parameter, which is shown in Fig.~\ref{fig:sentivity}, provides an intuition that the modification of the probability ratio makes no difference in the gradients of the vulnerability detection ratio curve. Even a larger increasing probability ratio may speed up the detection in the first period; the final estimated number of steps to detect all vulnerabilities is almost indistinguishable.

\begin{figure}
    \centering
    \includegraphics[width=0.5\textwidth]{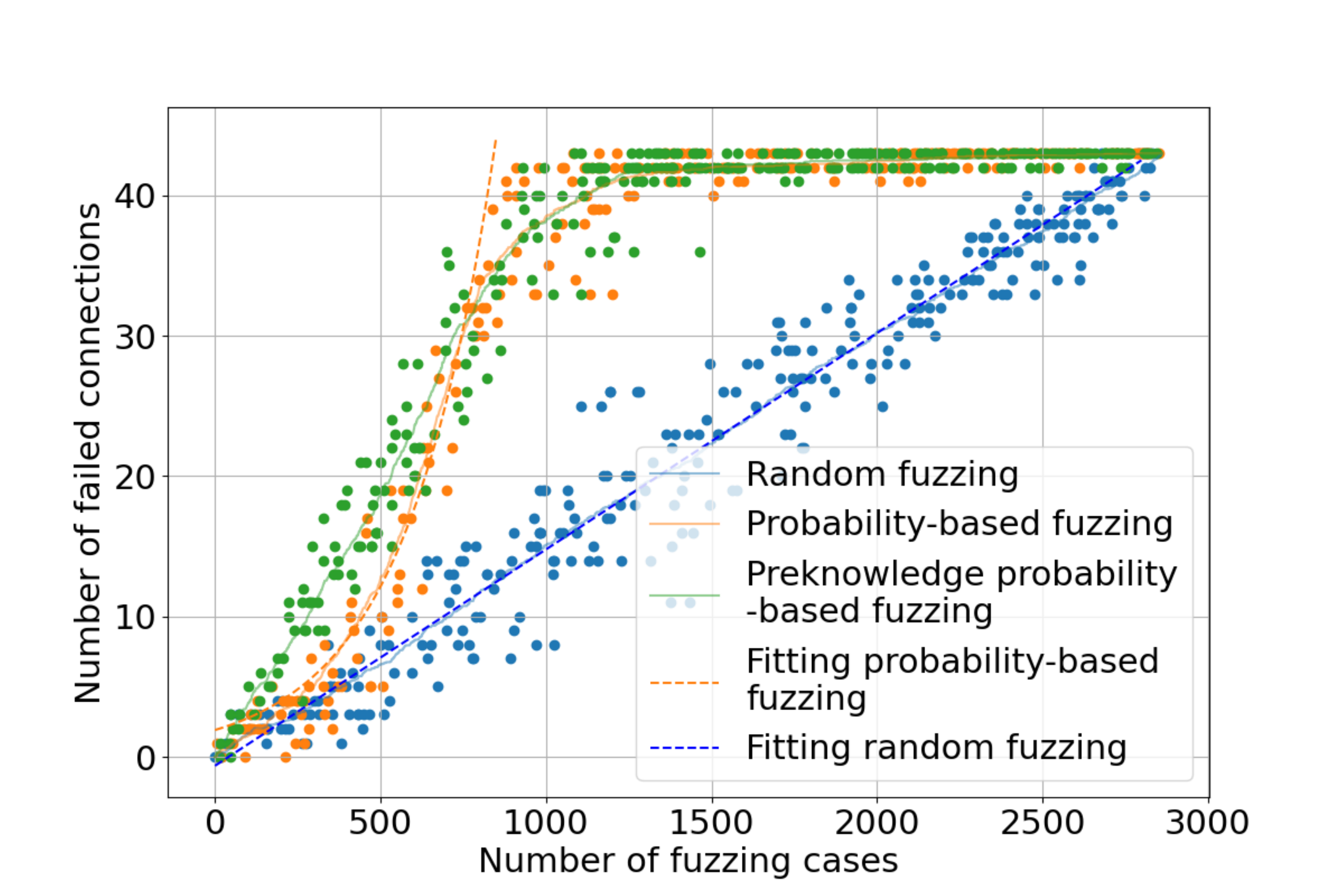}
    \caption{Comparison of Benchmark random-based fuzzing and probability-based fuzzing.}
    \label{fig:fuzzing_compare}
    \vspace{-15pt}
\end{figure}

We use several algorithms to fit strategies and generate the best representations for each strategy. With the assumption that the number of fuzzed cases is $i$, we find that the regress exponential algorithm, $2.072\times\mathrm{e}^{0.004i}$ orange dashed line in Fig.~\ref{fig:fuzzing_compare}, is the best fitting algorithm to the beginning period of probability-based fuzzing with $R^{2}$ value of 0.972. And for the end of probability-based fuzzing, vulnerabilities that do not follow the learned pattern are the primary reason for slow-down growth, like the case which changes command AV to command U shown as blue square in Fig.~\ref{fig:fuzzing}(b). Furthermore, for random fuzzing, the linear algorithm, $0.015i - 0.617$ blue dashed line in Fig.~\ref{fig:fuzzing_compare}, is the best fitting algorithm with $R^{2}$ value of 0.987.

To accelerate the speed of vulnerability detection in the beginning period, we design a probability-based fuzzing strategy with extra prior knowledge. In the preknowledge probability-based fuzzing strategy, we assign two arbitrary vulnerabilities as the extra prior knowledge to skip the pattern collection procedures in the beginning period. We run 20 times of each strategy, and plot the average and part of random points in Fig.~\ref{fig:fuzzing_compare}. The random fuzzing strategy has the worst performance among the three strategies, and the other strategies have similar performance except for the beginning period. In the beginning period, the prior knowledge can provide a local gaudiness for the probability fuzzer to efficiently locate the vulnerabilities, which is twice faster than the probability-based fuzzing strategy, especially in the first 500 fuzzing cases. Then, we can take advantage of extra prior knowledge to speed up the short-term efficiency of vulnerability detection.

To better understand the detected vulnerabilities in our proposed probability-based fuzzing strategy, we have selected some commands as samples, listed in Table III, to analyze and evaluate their impacts on both the RRC layer and the MAC layer. The results, as shown in Table IV, indicate that reactions vary between different layers. For instance, replacing command A with command B in the RRC layer will pass the integrity check at the gNB but lead to a failed connection due to the inconsistency caused by the command substitution. However, replacing Command A with Command B in the MAC layer will lead to repeated transmission of commands from the UE, effectively resulting in a Denial of Service (DoS) attack.

\begin{table}[]
\caption{Command List}
\label{tab:command_list}
\centering
\begin{tabular}{|c|l|}
\hline
Command Index & Command Type                  \\ \hline
A             & RRC connection request        \\ \hline
B             & RRC connection setup complete \\ \hline
C             & UL information transfer       \\ \hline
D             & Security mode complete        \\ \hline
E             & UE capability information     \\ \hline
\end{tabular}

\end{table}

\begin{table}[]
\caption{Comparison of Fuzz Strategy in Different Platforms}
\label{tab:fuzz_analysis}
\centering
\begin{tabular}{|m{3.2cm}|m{1.8cm}|m{2cm}|}
\hline
Fuzz Strategy (Up-Link)           & RRC layer & MAC layer \\ \hline
command A $\rightarrow$ command D & Failed connection    & Repeat of command A   \\ \hline
command B $\rightarrow$ command D & Failed connection    & Failed connection     \\ \hline
\end{tabular}
\end{table}


\section{\ac{SoAL} Bit-level Strategy}
Source-and-Learn (\ac{SoAL}) Bit-level Strategy provides a digital twin of some active attacks such as overshadowing~\cite{ludant2021sigunder}, which can change part of identifier values of communicated commands. In this strategy, we take two approaches, before-encryption and after-encryption, to represent two different scenarios, with domain knowledge and without domain knowledge. Compared to the traditional psychical overshadowing test, our proposed bit-level strategy achieves efficient vulnerability detection, which helps us fuzz more vulnerable cases and focus on the protocol.

\subsection{Risk Prioritized Fuzzing strategy}
Bit-level fuzzing is to randomly change the value of different identifiers in the specific command to generate different fuzzing cases. Following the guidance of command-level fuzzing, we can take more efficient bit-level fuzzing to locate vulnerabilities. For instance, based on the result of command-level fuzzing, we can first do bit-level fuzzing on these high-risk commands, e.g., command C and command H in Fig.~\ref{fig:fuzzing}. 

During the bit-level fuzzing procedure, we set a message detection and multi-lists of identifiers which cover the value range. For any specific message, the system takes a random value which has never been used to replace the identifier. There are two replacement strategies on bit-level fuzzing: before-encryption and after-encryption. Before-encryption approach is to change the identifier values before the protocol encryption while after-encryption approach is implemented in a reserved way. In this way, we can try all possible fuzzing cases to specific identifiers in special commands.

\subsection{Result Assessment of Bit-level Fuzzing}

Based on the result of Qpaque command-level and domain probability-based command-level fuzzing. As shown in Table~\ref{tab:bit_fuzzing}, three high-risk commands, `RRC Setup Request,' `RRC Reconfiguration,' and `RRC connection,' are selected as our bit-level fuzzing target. 

In the `RRC Setup Request' command, we fuzz it with both the before-encryption and after-encryption approaches. For the before-encryption approach, there were three identifiers: `ue-Identity', `Establishment Cause', and `spare'. Since the `spare' identifier never contains critical information and only occupies 1 bit, we only do bit fuzzing on `ue-Identity' and `Establishment Cause'. As shown in Table \ref{tab:bit_fuzzing}, any value for `ue-Identity' will not affect the connection. However, a different value of `Establishment Cause' can make connection transfer into different service types, as also mentioned in \cite{bitsikas2021don}. For example, if we change `Establishment Cause' from bit 0110 to bit 0000, UE can only request an emergency call. Moreover, with the after-encryption approach, all fuzzing cases lead to disconnection. This shows the integrity check for 5G protocol can identify whether the message is modified or not.

To further analyze the mechanism of the attack model across different layers, we use the `RRC Setup Request' command as an example, as detailed in Table~{\ref{tab:bit_comparison}}. In the MAC layer, all fuzzing attempts, regardless of the command identifier, result in failed connections. This outcome clearly indicates that an integrity check is crucial during the authentication process, as bit-level fuzzing without regeneration of the integrity checksum effectively simulates a Denial of Service (DoS) attack.

Except for the `RRC Setup Request' command, we also fuzz other two downlink commands, `RRC Reconfiguration' (command C in Fig. \ref{fig:fuzzing}) and `RRC Connection' (command H in Fig. \ref{fig:fuzzing}). In `RRC Reconfiguration' command, we take `sr-ConfigIndex' as our target because this identifier is responsible for radio scheduling and critical for connection establishment. However, no matter how we modify the `sr-ConfigIndex', the connection between \ac{UE} and gNB can still be established. On the contrary, when we fuzz the identifier `srb1\_srn\_id' in `RRC Connection' command with different values, UE rejects the connection establishment. Therefore, we conclude that UE may have alternative methods to negotiate the `sr-ConfigIndex', but cannot accept a new `srb\_id'. 

\begin{table}[]
\caption{Result of Bit-level Fuzzing}
\label{tab:bit_fuzzing}
\resizebox{\columnwidth}{!}{%
\begin{tabular}{|c|c|c|c|c|}
\hline
Channel & Command & Identifier & Fuzzing Value & Result                                                                    \\ \hline
\multirow{20}{*}{\begin{tabular}[c]{@{}c@{}}CCCH\\ (Up\_Link)\end{tabular}} &
  \multirow{20}{*}{\begin{tabular}[c]{@{}c@{}}RRC\\ Setup Request\end{tabular}} &
  \multirow{3}{*}{ue-Identity} &
  00 &
  successful \\ \cline{4-5} 
        &         &            & 01            & successful                                                                \\ \cline{4-5} 
        &         &            & 10            & successful                                                                \\ \cline{3-5} 
 &
   &
  \multirow{16}{*}{\begin{tabular}[c]{@{}c@{}}Establishment\\ Cause\end{tabular}} &
  0000 &
  emergency \\ \cline{4-5} 
        &         &            & 0001          & nulltype                                                                  \\ \cline{4-5} 
        &         &            & 0010          & high\_prio\_access                                                        \\ \cline{4-5} 
        &         &            & 0011          & nulltype                                                                  \\ \cline{4-5} 
        &         &            & 0100          & mt\_access                                                                \\ \cline{4-5} 
        &         &            & 0101          & nulltype                                                                  \\ \cline{4-5} 
        &         &            & 0110          & mo\_sig                                                                   \\ \cline{4-5} 
        &         &            & 0111          & nulltype                                                                  \\ \cline{4-5} 
        &         &            & 1000          & mo\_data                                                                  \\ \cline{4-5} 
        &         &            & 1001          & nulltype                                                                  \\ \cline{4-5} 
        &         &            & 1010          & \begin{tabular}[c]{@{}c@{}}delay\_tolerant\\ \_access\_v1020\end{tabular} \\ \cline{4-5} 
        &         &            & 1011          & nulltype                                                                  \\ \cline{4-5} 
        &         &            & 1100          & \begin{tabular}[c]{@{}c@{}}mo\_voice\\ \_call\_v1280\end{tabular}         \\ \cline{4-5} 
        &         &            & 1101          & nulltype                                                                  \\ \cline{4-5} 
        &         &            & 1110          & spare1                                                                    \\ \cline{4-5} 
        &         &            & 1111          & nulltype                                                                  \\ \cline{3-5} 
        &         & spare      & None          & None                                                                      \\ \hline
\multirow{7}{*}{\begin{tabular}[c]{@{}c@{}}PDCCH\\ (Down\_Link)\end{tabular}} &
  \multirow{7}{*}{\begin{tabular}[c]{@{}c@{}}RRC\\ Reconfiguration\end{tabular}} &
  \multirow{7}{*}{sr-Configindex} &
  \textless 5 &
  successful \\ \cline{4-5} 
        &         &            & \textless 15  & successful                                                                \\ \cline{4-5} 
        &         &            & \textless 35  & successful                                                                \\ \cline{4-5} 
        &         &            & \textless 75  & successful                                                                \\ \cline{4-5} 
        &         &            & \textless 155 & successful                                                                \\ \cline{4-5} 
        &         &            & \textless 157 & successful                                                                \\ \cline{4-5} 
        &         &            & = 157         & successful                                                                \\ \hline
\multirow{2}{*}{\begin{tabular}[c]{@{}c@{}}CCCH\\ (Down\_Link)\end{tabular}} &
  \multirow{2}{*}{\begin{tabular}[c]{@{}c@{}}RRC\\ Connection\end{tabular}} &
  \multirow{2}{*}{srb1.srb\_id} &
  0 &
  Reject \\ \cline{4-5} 
        &         &            & 2             & Reject                                                                    \\ \hline
\end{tabular}%
}
\end{table}

\begin{table}
    \centering
    \caption{Comparison of Different Bit-level Strategies}
    \resizebox{\columnwidth}{!}{%
\begin{tabular}{|c|c|c|c|c|}
\hline
Command &
  Identifier &
  \begin{tabular}[c]{@{}c@{}}Fuzzing \\ Value\end{tabular} &
  \begin{tabular}[c]{@{}c@{}}Fuzzing in \\ RRC Layer\end{tabular} &
  \begin{tabular}[c]{@{}c@{}}Fuzzing in \\ MAC Layer\end{tabular} \\ \hline
\multirow{19}{*}{\begin{tabular}[c]{@{}c@{}}RRC\\ Connection\\ Request\end{tabular}} &
  \multirow{3}{*}{UE-Identity} &
  00 &
  successful &
  failed \\ \cline{3-5} 
 &  & 01   & successful                                                                & failed \\ \cline{3-5} 
 &  & 10   & successful                                                                & failed \\ \cline{2-5} 
 &
  \multirow{16}{*}{\begin{tabular}[c]{@{}c@{}}Establishment\\ Cause\end{tabular}} &
  0000 &
  emergency &
  failed \\ \cline{3-5} 
 &  & 0001 & nulltype                                                                  & failed \\ \cline{3-5} 
 &  & 0010 & high\_prio\_access                                                        & failed \\ \cline{3-5} 
 &  & 0011 & nulltype                                                                  & failed \\ \cline{3-5} 
 &  & 0100 & mt\_access                                                                & failed \\ \cline{3-5} 
 &  & 0101 & nulltype                                                                  & failed \\ \cline{3-5} 
 &  & 0110 & mo\_sig                                                                   & failed \\ \cline{3-5} 
 &  & 0111 & nulltype                                                                  & failed \\ \cline{3-5} 
 &  & 1000 & mo\_data                                                                  & failed \\ \cline{3-5} 
 &  & 1001 & nulltype                                                                  & failed \\ \cline{3-5} 
 &  & 1010 & \begin{tabular}[c]{@{}c@{}}delay\_tolerant\\ \_access\_v1020\end{tabular} & failed \\ \cline{3-5} 
 &  & 1011 & nulltype                                                                  & failed \\ \cline{3-5} 
 &  & 1100 & \begin{tabular}[c]{@{}c@{}}mo\_voice\\ \_call\_v1280\end{tabular}         & failed \\ \cline{3-5} 
 &  & 1101 & nulltype                                                                  & failed \\ \cline{3-5} 
 &  & 1110 & spare1                                                                    & failed \\ \cline{3-5} 
 &  & 1111 & nulltype                                                                  & failed \\ \hline
\end{tabular}%
}
    
    \label{tab:bit_comparison}
\end{table}

\begin{figure}[!t]
    \centering
    
    \includegraphics[trim={2cm 0 2cm 0 },width=0.45\textwidth]{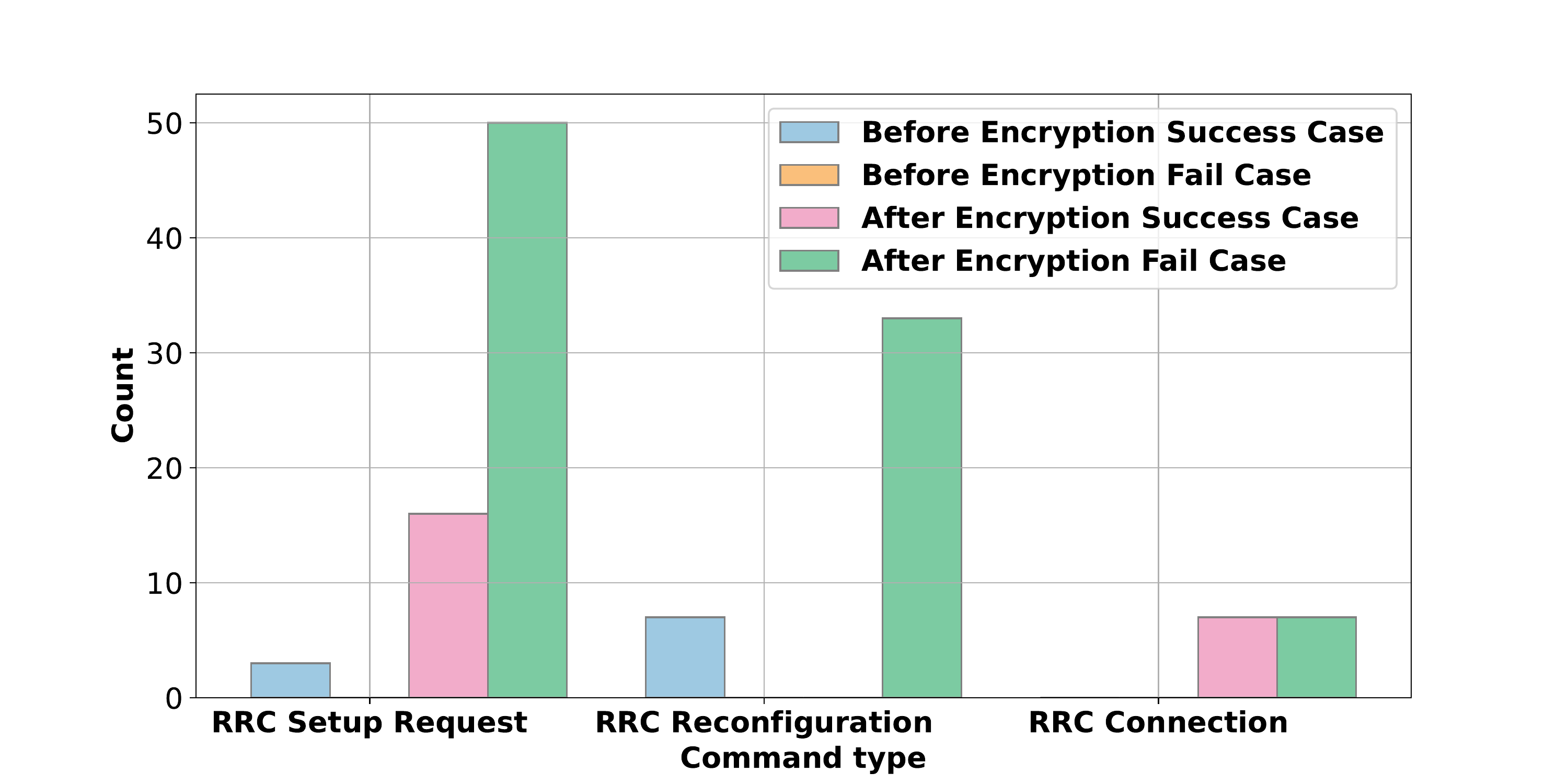}
    \caption{RRC high-risk command fuzzing distribution}
    \label{fig:bit_fuzzing_compare}
    \vspace{-15pt}
\end{figure}

In Fig. \ref{fig:bit_fuzzing_compare}, our system shows the ability to detect vulnerabilities efficiently. Among all 33 possible before-encryption cases, our system detects 10 vulnerabilities. As for the after-encryption approach, we find that it is unlikely to find a successful case because of the integrity check. Hence, we conclude that only fuzzing before-encryption is the only appropriate way for vulnerability detection in bit-level fuzzing. 

\section{Comparison Analysis}
As shown in Table{~\ref{tab:comparison_fuzz}}, in the landscape of fuzzing methodologies, our models (LAL, SyAL, SoAL) emerge as a sophisticated evolution from existing paradigms represented by BASESAFE{~\cite{maier2020basesafe}}, BaseSPEC{~\cite{kim2021basespec}}, FIRMWIRE{~\cite{hernandez2022firmwire}}, LTEFuzz{~\cite{kim2019touching}}, LTEInspector{~\cite{hussain2018lteinspector}},
and Directfuzz{~\cite{canakci2021directfuzz}}. Each model is characterized by a specific approach that dictates the required level of pre-knowledge and the degree of automation in the fuzzing process.
Adopting a black box approach, the LAL model eliminates the requirements for complex system knowledge and gets the capacity of automatically command-level fuzzing. This model enables an extensive evaluation of the system’s external interfaces, independent of the internal architecture. This black box approach contrasts sharply with the semi-automated, white box strategies of models like BASESAFE and BaseSPEC, which require comprehensive internal system understanding for testing baseband and conducting non-task boundary fuzzing, respectively.
On the other hand, the SyAL model is categorized within the grey box domain, denoting a moderate level of protocol structure comprehension, thereby refining the precision of the fully automated bit-level fuzzing capabilities. This model offers a compromise, bridging the extensive system insight required by white box models—such as LTEInspector, which engages in rule-based, command-level fuzzing with full access to LTE source software—and the no-prior-knowledge approach embodied by LAL.
Lastly, the SoAL model parallels the white box strategies by employing an exhaustive knowledge-based, semi-automated strategy with bit-level fuzzing. Compared to FIRMWIRE’s prerequisite of baseband initialization through boot procedures, this approach can thoroughly examine the system’s vulnerabilities without root privilege.
We collectively present various models with different automation levels and knowledge prerequisites, enabling them to be flexibly deployed across various testing scenarios. The LAL model is notable for its operational independence from system pre-knowledge; SyAL strikes a balance with limited insight into the protocol structure; and SoAL aligns with traditional white box methods in its detailed approach but surpasses them in automation efficiency.
\begin{table*}[]
\centering
\caption{Comparison Analysis with Different Fuzzing Strategies}
\label{tab:comparison_fuzz}
\begin{tabular}{|cc|c|c|c|l|}
\hline
\multicolumn{2}{|c|}{Approaches} &
  System Model &
  \begin{tabular}[c]{@{}c@{}}Prior-knowledge\\Requirement
  \end{tabular}&
  Fuzzing Technologies &
  Automation Level \\ \hline
\multicolumn{2}{|c|}{BASESAFE} &
  White box &
  Baseband firmware&
 \begin{tabular}[c]{@{}c@{}}Command-level fuzzing in \\ 
 non-task boundaries\end{tabular}&
  Semi-automated \\ \hline
\multicolumn{2}{|c|}{BaseSPEC} &
  White box &
  Baseband firmware &
   \begin{tabular}[c]{@{}c@{}}Bit-level fuzzing in \\ 
 non-task boundaries\end{tabular} &
  Semi-automated \\ \hline
\multicolumn{2}{|c|}{FIRMWIRE} &
  White box &
  \begin{tabular}[c]{@{}c@{}}Initialize the baseband\\ through boot\end{tabular} &
  \begin{tabular}[c]{@{}c@{}}Command-level fuzzing in \\ 
 Baseband’s RTOS APIs\end{tabular}&
  Semi-automated \\ \hline
\multicolumn{2}{|c|}{LTEFuzz} &
  White box &
  \begin{tabular}[c]{@{}c@{}}Fully controllable LTE\\open source software\end{tabular} &
  \begin{tabular}[c]{@{}c@{}}Rule-based \\ 
  command-level fuzzing\end{tabular}&
  Semi-automated \\ \hline
\multicolumn{2}{|c|}{LTEInspector} &
  White box &
  \begin{tabular}[c]{@{}c@{}}Abstract LTE model\end{tabular} &
  \begin{tabular}[c]{@{}c@{}}Rule-based \\ 
  command-level fuzzing\end{tabular} &
  Semi-automated \\ \hline
\multicolumn{2}{|c|}{Directfuzz} &
  Gray box &
  \begin{tabular}[c]{@{}c@{}}RTL input range\end{tabular} &
  \begin{tabular}[c]{@{}c@{}}Coverage guided \\ 
  bit-level fuzzing\end{tabular} &
  Fully-automated \\ \hline
\multicolumn{1}{|c|}{\multirow{3}{*}{Ours}} &
  \begin{tabular}[c]{@{}c@{}}Listen-and-Learn\\ (LAL)\end{tabular} &
  Black box &
  \begin{tabular}[c]{@{}c@{}}No prior-knowledge\\ required \end{tabular}&
  Command-level fuzzing &
  Fully-automated \\ \cline{2-6} 
\multicolumn{1}{|c|}{} &
  \begin{tabular}[c]{@{}c@{}}Synchronize-and\\ -Learn (SyAL)\end{tabular} &
  \begin{tabular}[c]{@{}c@{}}Gray box\\ (Protocol\\ Structure)\end{tabular} &
  Meassge inerpreter &
  \begin{tabular}[c]{@{}c@{}}Bit-level fuzzing\end{tabular} &
  Fully-automated \\ \cline{2-6} 
\multicolumn{1}{|c|}{} &
  \begin{tabular}[c]{@{}c@{}}Source-and-Learn\\ (SoAL)\end{tabular} &
  White box &
  SrsRAN modification &
  Bit-level fuzzing &
  Semi-automated \\ \hline
\end{tabular}

\end{table*}
\section{Conclusion}\label{conclusion}

We designed a novel fuzzing approach that systematically and cumulatively detects vulnerabilities for 5G and NextG protocols. Our approach is designed to detect, characterize, predict, and reveal vulnerabilities with varying levels of prior knowledge assumptions for attackers, beginning from no prior knowledge to full source code access. A digital twin framework was proposed and three fuzzing strategies were developed: LAL (black-box fuzzing), SyAL (gray-box fuzzing), and SoAL (white-box fuzzing). In black-box scenarios where no prior knowledge of the platform is known, the LAL strategy randomly arranges the sequence of commands. In gray-box scenarios where partial access to information is allowed, the SyAL strategy randomly replays the record commands with access to critical synchronization information for potential user information collection, which is similar to \ac{RNTI}. When the system is a white-box that supports full access to source code, the SoAL method performs bit-level fuzzing guided by the command-level fuzzing of LAL and SyAL for risk analysis transparency and reasoning.

In particular, the LAL strategy detected 129 vulnerabilities with 39 command types using only transmitted messages, and the embedded LSTM model efficiently predicted over 89\% of connection failures in 0.072 seconds on average. We then proposed a probability-based vulnerability detection method in the SyAL strategy, which achieves a linear growth of time cost with system size and allows for the detection of all vulnerabilities with partial user privacy information. This outperforms traditional fuzzing models with exponential growth of time consumption. In addition, based on the results of the SyAL strategy, the proposed SoAL method not only validates the integrity mechanism of 5G protocols but also detects three types of man-in-the-middle (MITM) vulnerabilities that are critical to protocol security. Extensive simulation results demonstrated that the designed fuzzing system is an efficient, automated approach that supports real-time vulnerability detection and proactive defense for existing 5G platforms and future released protocols.


\section*{Acknowledgment}
This effort was sponsored by the Defense Advanced Research Project Agency (DARPA) under grant no. D22AP00144. The views and conclusions contained herein are those of the authors and should not be interpreted as necessarily representing the official policies or endorsements, either expressed or implied, of DARPA or the U.S. Government.

\bibliographystyle{IEEEtran}
\bibliography{reference,references_ying, references_ying_mendeley}

\begin{IEEEbiography}[{\includegraphics[width=1in,height=1.25in,clip,keepaspectratio]{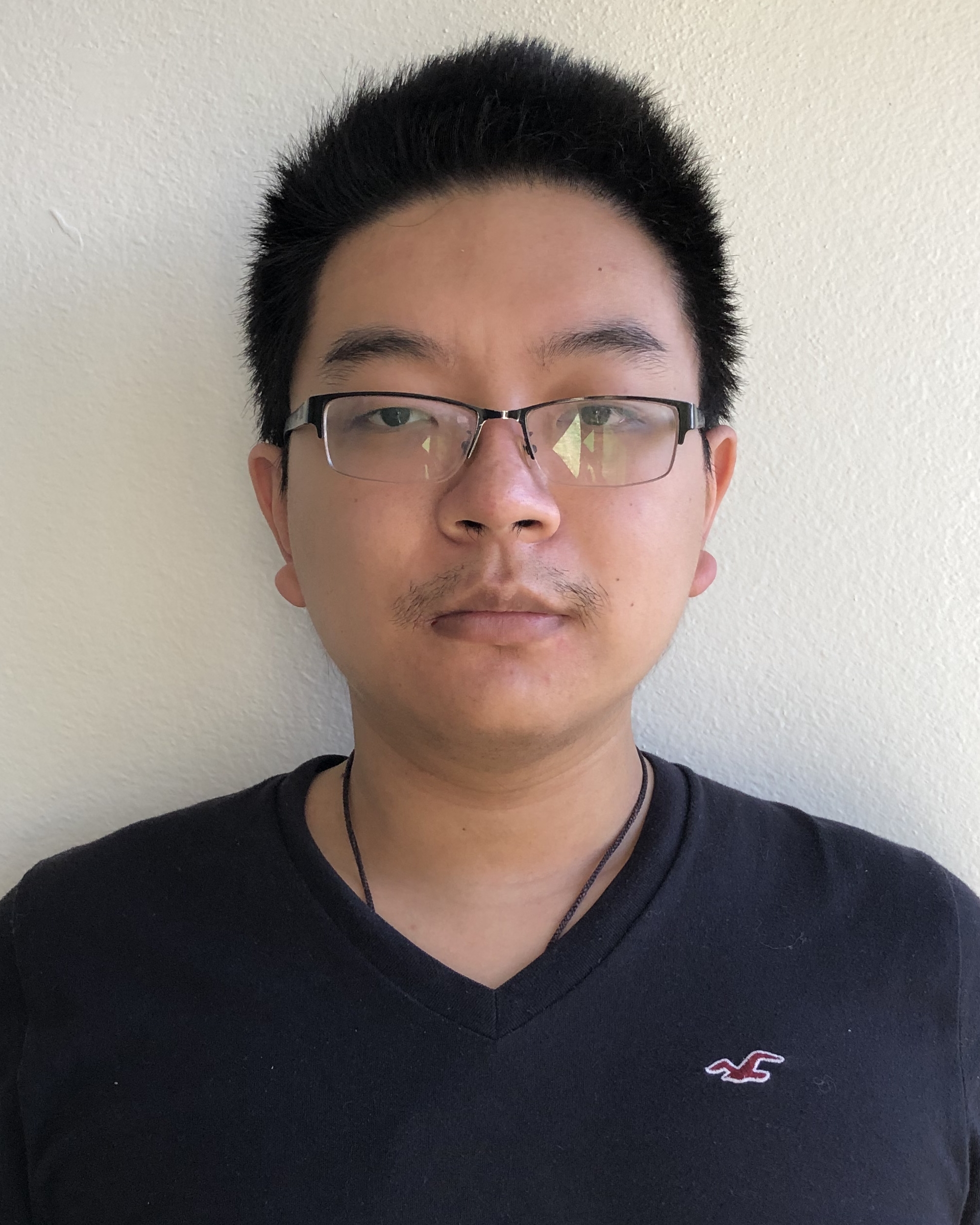}}]{Jingda Yang} (Graduate Student Member, IEEE) received the B.E. degree in software engineering from Shandong University and the M.Sc. degree in computer science from The George Washington University. He is currently a Ph.D. student in the School of System and Enterprises at Stevens Institute of Technology.  His research interests are formal verification and vulnerability detection of wireless protocol in 5G.
\end{IEEEbiography}

\begin{IEEEbiography}[{\includegraphics[width=1in,height=1.25in,clip,keepaspectratio]{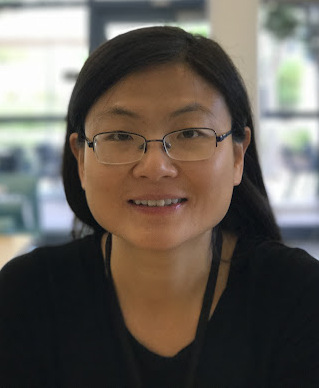}}]{Ying Wang} (Member, IEEE) received the B.E. degree in information engineering at Beijing University of Posts and Telecommunications, M.S. degree in electrical engineering from University of Cincinnati and the Ph.D. degree in electrical engineering from Virginia Polytechnic Institute and State University. She is an associate professor in the School of System and Enterprises at Stevens Institute of Technology. Her research areas include cybersecurity, wireless AI, edge computing, health informatics, and software engineering. 
\end{IEEEbiography}

\begin{IEEEbiography}[{\includegraphics[width=1in,height=1.25in,clip,keepaspectratio]{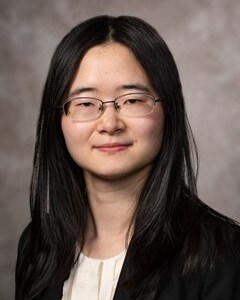}}]{Yanjun Pan} (Member, IEEE) received the B.E. degree in Information Engineering from Nanjing University of Aeronautics and Astronautics, China in 2016. She earned her M.S. degree and Ph.D. in Electrical and Computer Engineering from the University of Arizona, USA, in 2018 and 2021, respectively. She is an assistant professor in the Department of Computer Science and Computer Engineering at the University of Arkansas. Her current research interests include wireless security, information security, wireless sensing, and network optimization, with emphases on physical layer security, privacy-preserving technologies, mmWave sensing, and cross-layer optimization.
\end{IEEEbiography}

\begin{IEEEbiography}[{\includegraphics[width=1in,height=1.25in,clip,keepaspectratio]{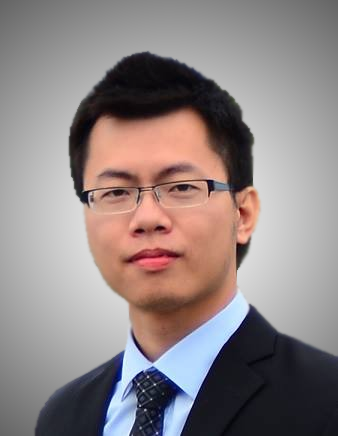}}]{Tuyen X. Tran} (Senior Member, IEEE) received the BEng degree (honors program) in Electronics and Telecommunications from the Hanoi University of Science and Technology, Vietnam, in 2011, the MSc degree in Electrical and Computer Engineering (ECE) from the University of Akron, Ohio, in 2013, and the PhD degree in ECE from Rutgers University, New Jersey, in 2018. His research interests include wireless communications, mobile cloud computing, and network optimization. He is currently a principal inventive scientist with AT\&T Labs Research, Bedminster, New Jersey.
\end{IEEEbiography}

\end{document}